\documentclass{JHEP3}

\usepackage{amsmath,amstext,amssymb}
\usepackage{psfrag,graphicx,subfigure}
\usepackage{cite}
 
\def\caln{{\cal N}}
\def\Str{{\text{Str}}}
\def\Fc{{\check F }}
\def\G{{\cal G}}
\def\H{{\cal H}}
\def\calo{{\cal O}}
\def\calc{{\cal C}}
\def\calt{{\cal T}}
\def\ie{{i.\,e.\ }}
\def\eg{{e.\,g.\ }}
\def\dt{\tilde d}
\def\mut{\tilde\mu}
\def\ft{\tilde f}
\def\vrho{\varrho}
\def\dt{\tilde d}
\def\At{\tilde A}
\def\pt{\tilde p}
\def\w{\mathfrak{w}}
\def\m{\mathfrak{m}}

\def\calw{{\cal W}}
\def\calf{{\cal F}}
\def\call{{\cal L}}
\def\re{\text{Re}\:}
\def\im{\text{Im}\:}

\def\del{\partial}

\def\str{\mathrm{Str}}
\def\ee{{\mathrm e}}
\def\ii{{\mathrm i}}

\newcommand{\dd}{\mathrm{d}}

\def\Z{{\mathbb Z}}

\title{Flavor Superconductivity from Gauge/Gravity Duality}

\author{Martin Ammon, Johanna Erdmenger, Patrick Kerner \\ Max-Planck-Institut f\"ur Physik (Werner-Heisenberg-Institut)\\
  F\"ohringer Ring 6, 80805 M\"unchen, Germany\\ \email{ammon, jke, pkerner@mppmu.mpg.de}}

\author{Matthias Kaminski\\ Instituto de Fisica Te\'orica UAM/CSIC  Facultad
  de Ciencias, C-XVI Universidad Aut\'onoma de Madrid Cantoblanco, Madrid
  28049, Spain\\ \email{matthias.kaminski@uam.es}}

\abstract{We give a detailed account and extensions of a holographic flavor
superconductivity model which we have proposed recently. The model has
an explicit field theory realization as strongly coupled $\mathcal{N}=2$
Super Yang-Mills theory with fundamental matter at finite temperature
and finite isospin chemical potential. Using gauge/gravity duality, \ie
a probe of two flavor D7--branes in the AdS black hole background, we
show that the system undergoes a second order phase transition with
critical exponent 1/2. The new ground state may be interpreted as a
$\rho$ meson superfluid. It shows signatures known from
superconductivity, such as an infinite dc conductivity and a gap in the
frequency-dependent conductivity. We present a stringy picture of the
condensation mechanism in terms of a recombination of strings. We give a
detailed account of the evaluation of the non-Abelian Dirac-Born-Infeld
action involved using two different methods. Finally we also consider
the case of massive flavors and discuss the holographic
Meissner--Ochsenfeld effect in our scenario. 
}

\date{\today}


\preprint{MPP-2009-29\\ IFT-UAM/CSIC-09-12}

\begin{document}

\section{Introduction and summary}
\label{sec:introduction-summary}

Gauge/gravity duality has been used very successfully over the past
years to obtain models with properties very similar to experimentally
observed phenomena in both elementary and condensed matter physics. 
This is due to the fact that
gauge/gravity duality is a powerful tool to compute 
observables of field theories at strong coupling \cite{Maldacena:1997re}. In view of
applications, it is particularly useful to turn on a temperature
using an AdS black hole as
gravity dual, as first proposed in \cite{Witten:1998zw}. 
In particular, gauge/gravity duality models which exhibit properties 
known from
superconductivity have recently been constructed \cite{Hartnoll:2008vx,
Nakano:2008xc,Albash:2008eh,Wen:2008pb,Gubser:2008wv,Roberts:2008ns,
Maeda:2008ir,Herzog:2008he,Basu:2008st,Horowitz:2008bn,Hartnoll:2008kx,Ammon:2008fc,
Basu:2008bh,Gubser:2008pf,O'Bannon:2008bz,Rebhan:2008ur,Evans:2008nf,
Denef:2009tp,Koutsoumbas:2009pa,Herzog:2009ci}.

In this context we have recently proposed a model which represents a
stringy realization of superconductivity in a relativistic framework \cite{Ammon:2008fc}. 
The aim of the present paper is to give a detailed account
of the setup and calculations involved in obtaining the results of
\cite{Ammon:2008fc}, as well as further new results.

Our model is based on embedding a probe of two coincident
D7 branes into
the AdS-black hole background, which extends
the original correspondence
to a field theory with flavor degrees of freedom in the fundamental
representation of the gauge group, along the lines of
\cite{Karch:2002sh, Kruczenski:2003be, Babington:2003vm}. Using this approach,
the correspondence has 
also been applied to finite flavor charge densities
\cite{Kobayashi:2006sb,Karch:2007br,Ghoroku:2007re,
Mateos:2007vc,Nakamura:2006xk,Nakamura:2007nx,Erdmenger:2008yj}. 
A finite charge density for the flavor
symmetry is obtained by giving a vev to the time
component of the gauge field on the D7 brane probe.
In~\cite{Erdmenger:2008yj}, a finite $SU(2)$ 
isospin density $\tilde d$ 
was considered using a probe of two coincident D7 branes and
switching on a gauge field $A=A_0^3 \dd x_0 \sigma^3$ on the
probe. At high isospin densities $\tilde d>\tilde d_c$, 
this configuration becomes unstable against 
gauge field fluctuations $X=a^1+\ii a^2$ and $Y=a^1-\ii a^2$, corresponding 
to vector mesons~\cite{Erdmenger:2007ja} 
in the field theory (see Figure 1) (see \cite{Apreda:2005yz} for an early
discussion of a similar instability). Note that we use capital letters for the
background and small letters for the fluctuations. 
\begin{figure}
  \centering
  \includegraphics[width=0.6\linewidth]{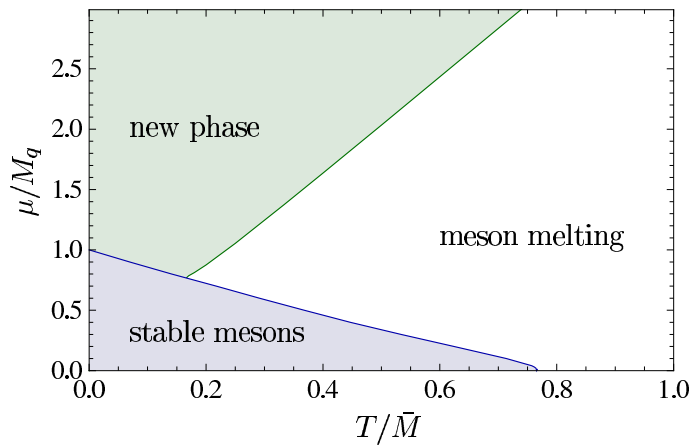}
  \caption{Phase diagram for fundamental matter in thermal strongly-coupled
    $\caln=2$ SYM theory~\cite{Erdmenger:2008yj}, with $\mu$ the
isospin chemical potential, $M_q$ the bare quark mass, $\bar M = 2 M_q \lambda^{-1/2}$,  $\lambda$ the 't Hooft coupling and $T$ the temperature: 
In the blue shaded region, mesons are stable. 
In the white and green regions, the mesons melt. Here the new phase is
stabilized while it was unstable in~\cite{Erdmenger:2008yj}. In this phase we find some features known from superconductivity.}
  \label{fig:phasediagram}
\end{figure}

The central result obtained in this model \cite{Ammon:2008fc} 
is that there is a solution to 
the equation of motion for the field on the two flavor branes, 
which in addition to the finite isospin density has a further 
vev of the form  $A_3^1 \dd x_3 \sigma^1$. This solution has lower free energy
in the new phase shown in figure~1 and stabilizes this phase. This applies at
least to stability against the fluctuations $X$ and $Y$. Moreover we find
that this phase has properties of a superconductor, such as a second order
phase transition with a critical exponent of 1/2. The flavor current, which is
analogous to the electromagnetic current, displays infinite dc conductivity
and a gap in the frequency-dependent conductivity. Furthermore, there is a dual
string-theoretical picture of the Cooper pairs, as well as
dynamical generation of meson masses. 

The stabilization
mechanism encountered in this context is related to the results 
of~\cite{Buchel:2006aa}.  Within the Sakai-Sugimoto model, 
a vev of the form  $A_3^1 \dd x_3 \sigma^1$ was considered in 
\cite{Aharony:2007uu} in the context of $\rho$ meson condensation. 
To obtain a p-wave superconductor within a phenomenological AdS/QCD
model, a vev of the form  $A_3^1 \dd x_3 \sigma^1$ was first 
used in~\cite{Gubser:2008wv}. Note that in our model the equations 
of motion forbid a source term for $A_3^1,$ \ie the breaking of 
the $U(1)$ is spontaneous.

The results presented here can be interpreted in two different ways.
On the one hand, our results provide an important link between 
gauge/gravity with flavor and
the recent gauge/gravity realizations of superconducting states  \eg in
\cite{Gubser:2008wv,Hartnoll:2008vx,Hartnoll:2008kx}, as well as
a realization of a superconducting current from a top-down
(super-)gravity model, rather than in a phenomenological bottom-up AdS/QCD
model.  Moreover, the degrees of freedom in the dual field theory are
identified explicitly. 

On the other hand, our holographic description can also be
considered as a realization of a meson superfluid. In the model presented here,
the $\rho$ mesons form a superfluid. In QCD like theories, $\rho$ meson
condensation is discussed in \cite{Brown:1991kk,Sannino:2002wp}. The corresponding picture is that
the mesons move through the quark-gluon plasma without friction. 
Similarly, the absorption rate for radiation should be reduced for a
specific frequency range \cite{Mateos:2007yp}. For pions, similar results 
have been found within QCD: There are analytical results for small and very
large isospin chemical potential using the chiral Lagrangian
\cite{Son:2000xc,Son:2000by}, as well as lattice results \cite{Kogut:2002zg}.
Unlike the lattice results, the results presented here can easily be
generalized to finite baryon chemical potential. A pion superfluid was
obtained within gauge/gravity duality in \cite{Rebhan:2008ur}. 

Let us compare our results to QCD. In QCD, the pion condensate is of
course the natural state in isospin asymmetric matter. The condensation of a
particle sets in if the isospin 
chemical potential is larger than the mass of this particle. According
to this rule, the pions condense first in QCD since they are the
Nambu-Goldstone bosons of the spontaneous chiral symmetry breaking and 
therefore the lightest particles. However the dual field theory which we
consider in this letter is supersymmetric at zero temperature and
therefore chiral symmetry cannot be broken spontaneously. In this
supersymmetric theory, the vector and scalar mesons have the same mass
at zero temperature. Due to finite temperature effects, the mass of the
vector and scalar mesons can become different as we increase the
temperature. It is a priori unclear which particle will condense. In our model
we checked that the vector mesons condense first such that the $\vrho$-meson
condensation state, which we consider in this paper, is the physical ground
state of our system near the phase transition.

In this paper we explain in detail the method used for evaluating the
non-Abelian Dirac-Born-Infeld action involved. Moreover we give a
detailed account of our results for finite quark mass and the
derivation of the Meissner effect.
Results similar to ours have also been obtained in \cite{Basu:2008bh} shortly
after \cite{Ammon:2008fc}. However there the 
non-abelian DBI action is considered only to second order in
$\alpha'$. Some of the features discussed here are also present 
in \cite{Basu:2008bh}: a second order phase transition 
as well as a gap in the frequency-dependent conductivity. 
However, since higher order terms in $\alpha^\prime$ are not taken 
into account in \cite{Basu:2008bh}, these authors are not able to see
higher order excitations in the frequency-dependent 
conductivity. Here we identify these higher order peaks with 
meson resonances.

Moreover, we give a string-theoretical interpretation of the formation
of a superconducting state of lower free energy: Essentially the
condensation process corresponds to a recombination of strings. 
The detailed picture is as follows: A small isospin density
corresponds to strings stretching from the D7 branes to the horizon
\cite{Kobayashi:2006sb}, the {\it horizon strings}. These are located
close to the horizon. Increasing the isospin density, 
the horizon strings strongly charge one of the D7 branes positively 
near the horizon, and  the other negatively. At a critical value, 
it becomes energetically favorable for the horizon strings to 
recombine into D7-D7 strings
which are free to move into the bulk and therefore lower the charge
density near the horizon. The D7-D7 strings form the superconducting
condensate. Thus they provide a picture of strongly coupled holographic Cooper
pairs. Note that for the construction considered the non-Abelian structure generated
by the two D$7$-branes is essential since it provides a charged condensate.
The non-Abelian forces play an important role for stabilizing the condensate.

This paper is organized as follows:  We begin with describing the
field-theory action in section \ref{sec:fieldtheory}. We then explain
the dual supergravity setup in section \ref{sec:holographic-setup}. In
particular, we contrast and compare two different approaches to
evaluating the non-Abelian DBI action involved. These are an adapted 
symmetrized trace prescription and the expansion of the DBI action to fourth 
order. We frequently refer to these approaches as prescriptions.
We then explain the string-theoretical picture in detail in section
\ref{sec:string-theory-pict}. Then we describe the thermodynamics of
the phase transition in \ref{sec:thermo}, and in particular show that it
is second order with critical exponent 1/2. 
In \ref{sec:fluctuations} we evaluate the fluctuations, again in the
two different approaches for the non-abelian DBI action. We calculate
the frequency-dependent conductivity and show it displays the expected
gap. Moreover we check that while the normal phase was unstable 
with respect to the fluctuations in $X$ and $Y$, the superconducting
phase is now stable with respect to these fluctuations. Finally we
show in \ref{sec:meissner} that our system displays a
Meissner-Ochsenfeld effect. We conclude with a collection of remarks
and suggestions in \ref{sec:discussion}.

\section{Field Theory Interpretation}
\label{sec:fieldtheory}
In this paper we will focus on a 3+1 -- dimensional $\mathcal{N}=2$
supersymmetric $SU(N_c)$ Yang-Mills theory at temperature $T$, consisting of a
$\mathcal{N}=4$ gauge multiplet as well as $N_f$ massive $\mathcal{N}=2$
supersymmetric hypermultiplets $(\psi_i, \phi_i).$ The hypermultiplets give
rise to the flavor degrees transforming in the fundamental representation of
the gauge group. The action is  written down explicitly for instance in
\cite{Erdmenger:2007cm}. In particular, we work in the large $N_c$ limit with
$N_f \ll N_c$ at strong coupling, \ie with $\lambda \gg 1,$ where $\lambda =
g_{\text{YM}}^2 N_c$ is the 't Hooft coupling constant. In the following we will
consider only two flavors, \ie $N_f=2.$ The flavor degrees of freedom are
called $u$ and $d$. If the masses of the two flavor degrees are degenerate,
the theory has a global $U(2)$ flavor symmetry, whose overall $U(1)_B$
subgroup can be identified with the baryon number. 

In the following we will consider the theory at finite isospin chemical potential $\mu$, which is introduced as the source of the operator
\begin{equation}
  \label{eq:7}
  J^3_0\propto \bar\psi\sigma^3\gamma_0\psi+\phi\sigma^3\del_0\phi=n_u-n_d\,,  
\end{equation}
where $n_{u/d}$ is the charge density of the isospin fields,
$(\phi_u,\phi_d)=\phi$ and $(\psi_u,\psi_d)=\psi$. $\sigma^i$ are the Pauli matrices. A non-zero vev $\langle J^3_0\rangle$ introduces an isospin density as discussed in
\cite{Erdmenger:2008yj}. The isospin chemical potential $\mu$ explicitly breaks the $U(2)\simeq
U(1)_B\times SU(2)_I$ flavor symmetry down to $U(1)_B\times U(1)_3$, where
$U(1)_3$ is generated by the unbroken generator $\sigma^3$ of the $SU(2)_I$.
Under the $U(1)_3$ symmetry the fields with index $u$ and $d$ have positive
and negative charge, respectively. 

However, the theory is unstable at large isospin chemical potential
\cite{Erdmenger:2008yj}. The new phase is sketched in figure
\ref{fig:phasediagram}. We show in this paper (see also \cite{Ammon:2008fc}),
that the new phase is stabilized by a non--vanishing vacuum expectation value
of the current component 
\begin{equation}
  \label{eq:1}
  \begin{split}
    J^1_3\propto \bar\psi\sigma^1\gamma_3\psi+\phi\sigma^1\del_3\phi=\bar\psi_u\gamma_3 \psi_d+\bar \psi_d\gamma_3 \psi_u+\text{bosons} \, .
  \end{split}
 \end{equation}
This current component breaks both the $SO(3)$ rotational symmetry as well as 
the remaining Abelian $U(1)_3$ flavor symmetry spontaneously. The rotational $SO(3)$ is broken down 
to $SO(2)_3$, which is generated by rotations around the $x^3$ axis. Due to
the non--vanishing vev for $J^1_3,$ flavor charged vector mesons condense and
form a superfluid. Let us emphasize that we do not describe a color
superconductor on the field theory side, since the condensate is a gauge
singlet.

Moreover, our results have a condensed matter interpretation:
Our model can be considered as a holographic 
p--wave superconductor. 
The global
$U(1)_3$ in our model is the analog of the local $U(1)_{\text{em}}$
symmetry of 
electromagnetic interactions. So far in all holographic models of superconductors
the breaking of a global symmetry on the field theory side is considered. In our 
model, the current $J^3$ corresponds to the electric 
current $J_{\text{em}}.$ The condensate $\langle J^1_3\rangle$ breaks
the $U(1)_3$ 
spontaneously. Therefore it can be viewed as the superconducting
condensate, 
which is analogous to the Cooper pairs. Since the 
condensate $\langle J^1_3\rangle$ transforms as a vector under spatial 
rotations, it is a p--wave superconductor.
-- Strictly speaking, for a superconductor interpretation
it would be necessary to 
gauge the global $U(1)_3$ symmetry which is broken spontaneously in
our model. A spontaneously broken global symmetry corresponds to a
superfluid.   However,  many features of 
superconductivity do not depend on whether the 
$U(1)_3$ is gauged. One exception to this is the Meissner--Ochsenfeld effect. 
To generate the currents expelling the magnetic field, the $U(1)_3$
symmetry has to be gauged. This is discussed further below in section
VII.


\section{Holographic Setup}
\label{sec:holographic-setup}

\subsection{Background and brane configuration} 
\label{sec:backgr-brane-conf}
  
We consider asymptotically $AdS_5\times S^5$ space-time. The $AdS_5\times
S^5$ geometry is holographically dual to the $\caln=4$ Super Yang-Mills
theory with gauge group $SU(N_c)$. The dual description of a finite
temperature field theory is an AdS black hole. We use the coordinates of
\cite{Kobayashi:2006sb} to write the AdS black hole background in Minkowski
signature as 
\begin{equation}
  \label{eq:AdSmetric}
  ds^2=\frac{\vrho^2}{2R^2}\left(-\frac{f^2}{\ft}\dd
    t^2+\ft
    \dd\vec{x}^2\right)+\left(\frac{R}{\vrho}\right)^2(\dd\vrho^2+\vrho^2\dd\Omega_5^2)\,,
\end{equation}
with $\dd\Omega_5^2$ the metric of the unit 5-sphere and
\begin{equation}
  f(\vrho)=1-\frac{\vrho_H^4}{\vrho^4},\quad
  \ft(\vrho)=1+\frac{\vrho_H^4}{\vrho^4}\,, 
\end{equation}
where $R$ is the AdS radius, with
\begin{equation}
  R^4=4\pi g_s N_c\,{\alpha'}^2 = 2\lambda\,{\alpha'}^2\,.
\end{equation}
The temperature of the black hole given by \eqref{eq:AdSmetric} may be
determined by demanding regularity of the Euclidean section. It is given by
\begin{equation}
  T=\frac{\vrho_H}{\pi R^2}\,. 
\end{equation}
In the following we may use the dimensionless coordinate
$\rho=\vrho/\vrho_H$, which covers the range from the event horizon at
$\rho=1$ to the boundary of the AdS space at $\rho\to\infty$.

To include fundamental matter, we embed $N_f$ coinciding D$7$-branes into
the ten-dimensional space-time. These D$7$-branes host flavor gauge fields
$A_\mu$ with gauge group $U(N_f)$. To write down the DBI action for the
D$7$-branes, we introduce spherical coordinates $\{r,\Omega_3\}$ in the
4567-directions and polar coordinates $\{L,\phi\}$ in the 89-directions
\cite{Kobayashi:2006sb}. The angle between these two spaces is denoted by
$\theta$ ($0\le\theta\le\pi/2$). The six-dimensional space in the
$456789$-directions is given by
\begin{equation}
    \dd\varrho^2+\varrho^2\dd\Omega_5^2=\,\dd r^2+r^2\dd\Omega_3^2+\dd L^2+L^2\dd\phi^2=\,\dd\varrho^2+\varrho^2(\dd\theta^2+\cos^2\theta\dd\phi^2+\sin^2\theta\dd\Omega_3^2)\,,
\end{equation}
where $r=\varrho\sin\theta$, $\varrho^2=r^2+L^2$ and $L=\varrho\cos\theta$.
Due to the $SO(4)$ rotational symmetry in the 4567 directions, the embedding of the D$7$-branes only depends on the
radial coordinate $\rho$. Defining $\chi=\cos\theta$, we parametrize the
embedding by $\chi=\chi(\rho)$ and choose $\phi=0$ using the $SO(2)$
symmetry in the 89-direction. The induced metric $G$ on the D$7$-brane
probes is then
\begin{equation}
  \label{eq:inducedmetric}
    ds^2(G)=\frac{\vrho^2}{2R^2}\left(-\frac{f^2}{\ft}\dd
      t^2+\ft\dd\vec{x}^2\right)+\frac{R^2}{\vrho^2}\frac{1-\chi^2+\vrho^2(\del_\vrho\chi)^2}{1-\chi^2}\dd\vrho^2+R^2(1-\chi^2)\dd\Omega_3^2\,.
\end{equation}
The square root of the determinant of $G$ is given by
\begin{equation}
  \sqrt{-G}=\frac{\sqrt{h_3}}{4}\varrho^3f\ft(1-\chi^2)\sqrt{1-\chi^2+\varrho^2(\del_\varrho\chi)^2}\,,
\end{equation}
where $h_3$ is the determinant of the 3-sphere metric.

As in \cite{Erdmenger:2008yj} we introduce a $SU(2)$ isospin chemical
potential $\mu$ by a non-vanishing time component of the non-Abelian background field
on the D$7$-brane. The generators of the $SU(2)$ gauge group are given by the
Pauli matrices $\sigma^i$. Due to the gauge symmetry, we may rotate the flavor
coordinates until the chemical potential lies in the third flavor direction,
\begin{equation}
  \label{eq:isospinchempot}
  \mu=\lim_{\rho\to\infty}A_0^3(\rho)\,.
\end{equation}
This non-zero gauge field breaks the $SU(2)$ gauge symmetry down to $U(1)_3$
generated by the third Pauli matrix $\sigma^3$. The spacetime symmetry on the
boundary is still $SO(3)$. Notice that the Lorentz group $SO(3,1)$ is already broken down
to $SO(3)$ by the finite temperature. In addition, we consider a further
non-vanishing background gauge field which stabilizes the system for large
chemical potentials. Due to the symmetry of our setup we may choose $A_3^1\dd
x^3\sigma^1$ to be non-zero. To obtain an isotropic configuration in the field
theory, this new gauge field $A_3^1$ only depends on $\rho$. Due to this two
non-vanishing gauge fields, the field strength tensor on the branes has the
following non-zero components,
\begin{equation}
  \label{eq:nonzeroF}
  \begin{split}
    &F^1_{\vrho 3}=-F^1_{3\vrho}=\del_\vrho A^1_3\,,\\
    &F^2_{03}=-F^2_{30}=\frac{\gamma}{\sqrt{\lambda}}A^3_0A^1_3\,,\\
    &F^3_{\vrho 0}=-F^3_{0\vrho}=\del_\vrho A^3_0\,.
  \end{split}
\end{equation}

\subsection{DBI action and equations of motion}
\label{sec:dbi-action-equations}
In this section we calculate the equations of motion which determine the
profile of the D$7$-brane probes and of the gauge fields on these branes. An
discussion of the gauge field profiles, which we use to give a geometrical
interpretation of the stabilization of the system and the pairing mechanism, may
be found in section \ref{sec:string-theory-pict}.

The DBI action determines the shape of the brane embeddings, \ie the scalar
fields $\phi$, as well as the configuration of the gauge fields $A$ on these
branes. We consider the case of $N_f=2$ coincident D$7$-branes for which the
non-Abelian DBI action reads \cite{Myers:1999ps}
\begin{equation}
 \label{eq:non-AbelianDBI}
 S_{\text{DBI}}=-T_{D7}\:\str\int\!\dd^8\xi\:\sqrt{\det
     Q}\Bigg[\det\Big(P_{ab}\big[E_{\mu\nu}+E_{\mu i}(Q^{-1}-\delta)^{ij}E_{j\nu}\big]+2\pi\alpha'F_{ab}\Big)\Bigg]^{\frac{1}{2}}
\end{equation}
with
\begin{equation}
 \label{eq:defQDBI}
 Q^i{}_j=\delta^i{}_j+\ii 2\pi\alpha'[\Phi^i,\Phi^k]E_{kj}
\end{equation}
and $P_{ab}$ the pullback to the D$p$-brane, where for a D$p$-brane in $d$ dimensions we have 
$\mu,\,\nu=0,\dots, (d-1)$, $a,\,b=0,\dots, p$, $i,\,j = (p+1),\dots, (d-1)$,
$E_{\mu\nu} = g_{\mu\nu} + B_{\mu\nu}$. In our case we set $p=7$, $d=10$,
$B\equiv 0$. As in \cite{Erdmenger:2008yj} we can simplify this action
significantly by using the spatial and gauge symmetries present in our setup.
The action becomes
\begin{equation}
  \label{eq:DBI}
  \begin{split}
    S_{\text{DBI}}&=-T_{D7}\int\!\dd^8\xi\:\str\sqrt{|\det(G+2\pi\alpha'F)|}\\
    &=-T_{D7}\int\!\dd^8\xi\:\sqrt{-G}\:\str\Bigg[1+G^{00}G^{44}\left(F^3_{\vrho
        0}\right)^2\left(\sigma^3\right)^2+G^{33}G^{44}
    \left(F^1_{\vrho 3}\right)^2\left(\sigma^1\right)^2\\
    & +G^{00}G^{33}\left(F^2_{03}\right)^2\left(\sigma^2\right)^2\Bigg]^{\frac{1}{2}}\,,
  \end{split}
\end{equation}
where in the second line the determinant is calculated. Due to the
symmetric trace, all commutators between the matrices $\sigma^i$ vanish. It is
known that the symmetrized trace prescription in the DBI action is only valid
up to fourth order in $\alpha'$ \cite{Tseytlin:1997csa,Hashimoto:1997gm}. However the corrections to the higher order
terms are suppressed by $N_f^{-1}$ \cite{Constable:1999ac} (see also \cite{Myers:2008me}). Here we use two different approaches to
evaluate the non-Abelian DBI action \eqref{eq:DBI}. First, we modify the
symmetrized trace prescription by omitting the commutators of the generators
$\sigma^i$ and then setting $(\sigma^i)^2=1$ (see subsection
\ref{sec:change-str-prescr} below). This prescription makes the calculation of
the full DBI action feasible. Second, we expand the non-Abelian DBI
action to fourth order in the field strength $F$ (see subsection \ref{sec:expansion-dbi-action}).

\subsubsection{Adapted symmetrized trace prescription}
\label{sec:change-str-prescr}
Using the adapted symmetrized trace prescription defined above, the action becomes
 \begin{equation}
   \label{eq:DBIchangedstr}
   \begin{split}
     S_{\text{DBI}}=&-T_{D7}N_f\int\!\dd^8\xi\:\sqrt{-G}\Bigg[1+G^{00}G^{44}\left(F^3_{\vrho 0}\right)^2+G^{33}G^{44}\left(F^1_{\vrho 3}\right)^2+G^{00}G^{33}
     \left(F^2_{03}\right)^2\Bigg]^{\frac{1}{2}}\\
     =&-\frac{T_{D7}N_f}{4}\!\int\!\dd^8\xi\;\vrho^3f\ft(1-\chi^2)\Upsilon(\rho,\chi,\At)\,,
   \end{split}
 \end{equation}
with
\begin{equation}
  \label{eq:Upsilon}
  \begin{split}
    \Upsilon(\rho,\chi,\At)=\Bigg[&1-\chi^2+\rho^2(\del_\rho\chi)^2-\frac{2\ft}{f^2}(1-\chi^2)\left(\del_\rho
      \At^3_0\right)^2+\frac{2}{\ft}(1-\chi^2)\left(\del_\rho \At^1_3\right)^2\\
    &-\frac{2\gamma^2}{\pi^2\rho^4f^2}(1-\chi^2+\rho^2(\del_\rho\chi)^2)
    \left(\At^3_0\At^1_3\right)^2\Bigg]^{\frac{1}{2}}\,,
  \end{split}
\end{equation}
where the dimensionless quantities $\rho=\vrho/\vrho_H$ and
$\At=(2\pi\alpha')A/\vrho_H$ are used. To obtain first order equations of
motion for the gauge fields which are easier to solve numerically, we perform a
Legendre transformation. Similarly to
\cite{Kobayashi:2006sb,Erdmenger:2008yj} we calculate the electric displacement
$p_0^3$ and the magnetizing field $p_3^1$ which are given by the conjugate
momenta of the gauge fields $A_0^3$ and $A_3^1$,
\begin{equation}
  \label{eq:conmomenta}
  p_0^3=\frac{\delta S_{\text{DBI}}}{\delta(\del_\vrho A^3_0)}\,,\qquad
  p_3^1=\frac{\delta S_{\text{DBI}}}{\delta(\del_\vrho A_3^1)}\,.
\end{equation}
In contrast to
\cite{Kobayashi:2006sb,Karch:2007br,Mateos:2007vc,Erdmenger:2008yj}, the
conjugate momenta are not constant any more but depend on the radial coordinate
$\vrho$ due to the non-Abelian term $A_0^3A_3^1$ in the DBI action. For the
dimensionless momenta $\pt_0^3$ and $\pt_3^1$ defined as  
\begin{equation}
  \label{eq:pt}
  \pt=\frac{p}{2\pi\alpha'N_fT_{D7}\vrho_H^3}\,,
\end{equation}
we get
\begin{equation}
  \label{eq:conmomentadim}
  \pt_0^3=\frac{\rho^3\ft^2(1-\chi^2)^2\del_\rho\At_0^3}{2f\Upsilon(\rho,\chi,\At)}\,,\quad\pt_3^1=-\frac{\rho^3f(1-\chi^2)^2\del_\rho\At_3^1}{2\Upsilon(\rho,\chi,\At))}\,.
\end{equation}
Finally, the Legendre-transformed action is given by
\begin{equation}
  \label{eq:DBIlegendre}
  \begin{split}
    \tilde S_{\text{DBI}}&=S_{\text{DBI}}-\int\!\dd^8\xi\:
    \Bigg[
      \left(\del_\vrho A_0^3\right)\frac{\delta S_{\text{DBI}}}{\delta
        \left(\del_\vrho A_0^3\right)}+\left(\del_\vrho A_3^1\right)\frac{\delta S_{\text{DBI}}}{\delta\left(\del_\vrho A_3^1\right)}
    \Bigg]\\
    &=-\frac{T_{D7}N_f}{4}\int\!\dd^8\xi\:\vrho^3f\ft(1-\chi^2)\sqrt{1-\chi^2+\rho^2(\del_\rho\chi)^2}\;V(\rho,\chi,\At,\pt)\,,
  \end{split}
\end{equation}
with
\begin{equation}
  \label{eq:V}
  V(\rho,\chi,\At,\pt)=\Bigg[
  \left(1-\frac{2\gamma^2}{\pi^2\rho^4f^2}\left(\At_0^3\At_3^1\right)^2\right)\Bigg(1+\frac{8
    \left(\pt_0^3\right)^2}{\rho^6\ft^3(1-\chi^2)^3}-\frac{8
    \left(\pt_3^1\right)^2}{\rho^6\ft f^2(1-\chi^2)^3}\Bigg)\Bigg]^{\frac{1}{2}}\,.
\end{equation}
Then the first order equations of motion for the gauge fields and their conjugate momenta are
\begin{equation}
  \label{eq:eomgauge}
  \begin{split}
    \del_\rho\At_0^3&=\frac{2f\sqrt{1-\chi^2+\rho^2(\del_\rho\chi)^2}}{\rho^3\ft^2(1-\chi^2)^2}\pt_0^3W(\rho,\chi,\At,\pt)\, , \\
    \del_\rho\At_3^1&=-\frac{2\sqrt{1-\chi^2+\rho^2(\del_\rho\chi)^2}}{\rho^3f(1-\chi^2)^2}\pt_3^1W(\rho,\chi,\At,\pt)\, , \\
    \del_\rho\pt_0^3&=\frac{\ft(1-\chi^2)\sqrt{1-\chi^2+\rho^2(\del_\rho\chi)^2}c^2}{2\pi^2\rho
      fW(\rho,\chi,\At,\pt)}\left(\At_3^1\right)^2\At_0^3\, , \\
    \del_\rho\pt_3^1&=\frac{\ft(1-\chi^2)\sqrt{1-\chi^2+\rho^2(\del_\rho\chi)^2}c^2}{2\pi^2\rho
      fW(\rho,\chi,\At,\pt)}\left(\At_0^3\right)^2\At_3^1\,,
  \end{split}
\end{equation}
with
\begin{equation}
  \label{eq:W}
   W(\rho,\chi,\At,\pt)=\sqrt{\frac{1-\frac{2\gamma^2}{\pi^2\rho^4f^2}\left(\At_0^3\At_3^1\right)^2}{1+\frac{8
      \left(\pt_0^3\right)^2}{\rho^6\ft^3(1-\chi^2)^3}-\frac{8
      \left(\pt_3^1\right)^2}{\rho^6\ft
      f^2(1-\chi^2)^3}}}\,.
\end{equation}
For the embedding function $\chi$ we get the second order equation of motion

\begin{equation}
  \label{eq:eomchi}
  \begin{split}
    \del_\rho
    \left[\frac{\rho^5f\ft(1-\chi^2)(\del_\rho\chi)}{\sqrt{1-\chi^2+\rho^2(\del_\rho\chi)^2}}V\right]
    =&-\frac{\rho^3f\ft\chi}{\sqrt{1-\chi^2+\rho^2(\del_\rho\chi)^2}}\Bigg[
    \left[3\left(1-\chi^2\right)+2\rho^2(\del_\rho\chi)^2\right]V\\
    &-\frac{24\left(1-\chi^2+\rho^2(\del_\rho\chi)^2\right)}{\ft^3\rho^6\left(1-\chi^2\right)^3}W
    \left(\left(\pt_0^3\right)^2-\frac{\ft^2}{f^2}\left(\pt_3^1\right)\right)\Bigg]\,.
  \end{split}
\end{equation}
We solve the equations numerically and determine the solution by integrating 
the equations of motion
from the horizon at $\rho=1$ to the boundary at $\rho=\infty$. The initial
conditions may be determined by the asymptotic expansion of the gravity fields near the
horizon
\begin{equation}
  \label{eq:asymh}
  \begin{aligned}
    \At_0^3&= &\frac{c_0}{\sqrt{(1-\chi_0^2)^3+c_0^2}}(\rho-1)^2&+\calo\left((\rho-1)^3\right)& \, , \\
    \At_3^1&=b_0& &+\calo\left((\rho-1)^3\right)& \,  , \\
    \pt_0^3&=c_0&+\frac{\gamma^2b_0^2c_0}{8\pi^2}(\rho-1)^2&+\calo\left((\rho-1)^3\right)& \, , \\
    \pt_3^1&=& &+\calo\left((\rho-1)^3\right)& \, , \\
    \chi&=\chi_0&-\frac{3\chi_0(1-\chi_0^2)^3}{4[(1-\chi_0^2)^3+c_0^2]}(\rho-1)^2&+\calo\left((\rho-1)^3\right)&\, ,
  \end{aligned}
\end{equation}
where
the terms in the expansions 
are arranged according to their order in $\rho-1$.
For the numerical calculation we consider the terms up to sixth order in $\rho - 1$. The
three independent parameter $b_0$, $c_0$ and $\chi_0$ may be determined by
field theory quantities defined via the asymptotic expansion of the gravity
fields near the boundary,
\begin{equation}
  \label{eq:asymb}
  \begin{aligned}
    \At_0^3&=\mut& &-\frac{\dt_0^3}{\rho^2}& &+\calo\left(\rho^{-4}\right)& \, , \\
    \At_3^1&=& &-\frac{\dt_3^1}{\rho^2} & &+\calo\left(\rho^{-4}\right)& \, , \\
    \pt_0^3&=\dt_0^3& & & &+\calo\left(\rho^{-4}\right)& \, , \\
    \pt_3^1&=-\dt_3^1& &+\frac{\gamma^2 \mut^2\dt_3^1}{4 \pi^2\rho^2} & &+\calo\left(\rho^{-4}\right)& \, , \\
    \chi&= &\frac{m}{\rho}&&+\frac{c}{\rho^3}&+\calo\left(\rho^{-4}\right)&\,.
  \end{aligned}
\end{equation}
According to the AdS/CFT dictionary, $\mu$ is the isospin chemical potential.
The parameters $\dt$ are related to the vev of the flavor currents $J$ by
\begin{equation}
  \label{eq:dt}
  \dt_0^3=\frac{2^{\frac{5}{2}}\langle
    J_0^3\rangle}{N_fN_c\sqrt{\lambda}T^3}\,,\quad \dt_3^1=\frac{2^{\frac{5}{2}}\langle J_3^1\rangle}{N_fN_c\sqrt{\lambda}T^3}
\end{equation}
and $m$ and $c$ to the bare quark mass $M_q$ and the quark condensate
$\langle\bar\psi\psi\rangle$, 
\begin{equation}
  \label{eq:cm}
  m=\frac{2M_q}{\sqrt{\lambda}T}\,,\quad c=-\frac{8\langle\bar\psi\psi\rangle}{\sqrt{\lambda}N_fN_cT^3}\, , 
\end{equation}
respectively. There are two independent physical parameters, \eg $m$ and $\mu$, in the grand
canonical ensemble. From the boundary asymptotics \eqref{eq:asymb}, we also
obtain that there is no source term for the current $J_3^1$. Therefore
as a non-trivial result we find that the
$U(1)_3$ symmetry is always broken spontaneously. In contrast, 
in the related works on
p-wave superconductors in $2+1$ dimensions
\cite{Gubser:2008wv,Roberts:2008ns}, the spontaneous breaking of the $U(1)_3$
symmetry has to be put in by hand by setting the source term for the
corresponding operator to zero. With the constraint $\At_3^1|_{\rho\to\infty}=0$ and the two
independent physical parameters, we may fix the three independent parameters of
the near-horizon asymptotics and obtain a solution to the equations of motion.

\subsubsection{Expansion of the DBI action}
\label{sec:expansion-dbi-action}
We now outline the second approach which we use. Expanding the action
\eqref{eq:DBI} to fourth order in the field strength $F$ yields
 \begin{equation}
   \label{eq:DBIaxpand}
   S_{\text{DBI}}=-T_{D7}N_f\int\!\dd^8\xi\:\sqrt{-G}
   \left[1+\frac{\calt_2}{2}-\frac{\calt_4}{8}+\cdots\right]\,,
 \end{equation}
where $\calt_i$ consists of the terms with order $i$ in $F$. To
calculate the $\calt_i$, we use the following results for the symmetrized traces
\begin{equation}
  \label{eq:str}
  \begin{aligned}
    \,&2\sigma:\quad &&\str\left[\left(\sigma^i\right)^2\right]=N_f \, , \\
    \,&4\sigma: &&\str\left[\left(\sigma^i\right)^4\right]=N_f\,,\quad\str\left[\left(\sigma^i\right)^2\left(\sigma^j\right)^2\right]=\frac{N_f}{3}\,,\\
  \end{aligned}
\end{equation}
where the indices $i,j$ are distinct. Notice that the symmetric trace of
terms with unpaired $\sigma$ matrices vanish, \eg
$\str[\sigma^i\sigma^j]=N_f\delta^{ij}$. The $\calt_i$ are given in the
appendix \ref{sec:expans-fourth-order}.

To perform the Legendre transformation of the above action, we determine the
conjugate momenta as in \eqref{eq:conmomenta}. However, we cannot easily solve these
equations for the derivative of the gauge fields since we obtain two coupled equations
of third degree. Thus we directly calculate the equations of motion for the
gauge fields on the D$7$-branes. The equations are given in
\eqref{eq:eomexpand} of the appendix.

To solve these equations, we use the same strategy as in the adapted symmetrized trace 
prescription discussed above. We integrate the equations of motion from the horizon at
$\rho=1$ to the boundary at $\rho=\infty$ numerically. The initial conditions
may be determined by the asymptotic behavior of the gravity fields near the
horizon
\begin{equation}
  \label{eq:asymhexp}
  \begin{aligned}
    \,&\At_0^3=& &a_2(\rho-1)^2&+\calo\left((\rho-1)^3\right)\,,\\
    \,&\At_3^1=&b_0&&+\calo\left((\rho-1)^3\right)\,,\\
    \,&\chi=&\chi_0&+\frac{3 (a_2^4+4a_2^2-8)\chi_0}{4 (3 a_2^4+4 a_2^2+8)}(\rho-1)^2&+\calo\left((\rho-1)^3\right)\,.
  \end{aligned}
\end{equation}
For the numerical calculation we use the asymptotic expansion up to sixth order.
As in the adapted symmetrized trace prescription, there are again three independent parameters
$a_2,b_0,\chi_0$. Since we have not performed a Legendre transformation, we
trade the independent parameter $c_0$ in the asymptotics of the conjugate
momenta $\pt_0^3$ in the symmetrized trace prescription with the independent parameter
$a_2$ (cf. asymptotics in equation \eqref{eq:asymh}). However, the
three independent parameters may again be determined in field theory quantities
which are defined by the asymptotics of the gravity fields near the boundary
\begin{equation}
  \label{eq:asymbexp}
  \begin{aligned}
    \,&\At_0^3=&\mu & &-\frac{\dt_0^3}{\rho^2}& &+\calo\left(\rho^4\right)\,,\\
    \,&\At_3^1=& & &-\frac{\dt_3^1}{\rho^2}& &+\calo\left(\rho^4\right)\,,\\
    \,&\chi=& &\frac{m}{\rho}&&+\frac{c}{\rho^3}&+\calo\left(\rho^4\right)\,.
  \end{aligned}
\end{equation}
The independent parameters $\mu,\dt_0^3,\dt_3^1,m,c$ are given by field theory
quantities as presented in \eqref{eq:dt} and \eqref{eq:cm}. Again we find that
there is no source term for the current $J_3^1$, which implies
spontaneous
symmetry breaking.  Therefore the independent
parameters in both prescriptions are the same and we can use the same strategy
to solve the equations of motion as described below \eqref{eq:cm}. 

\section{String Theory Picture}
\label{sec:string-theory-pict}
In this section we give a string theory interpretation, \ie a geometrical
picture, of the formation of a new phase, for which the field theory is 
discussed in section \ref{sec:fieldtheory}. We show that the system is stabilized by 
dynamically generating a non-zero vev of the current component
$J_3^1$ dual to the gauge field $A_3^1$ on the brane. 
Moreover, we find a geometrical picture of the pairing mechanism which
forms the condensate $\langle J_3^1\rangle$, the Cooper pairs.
\begin{figure}
  \centering
  \includegraphics[width=0.6\linewidth]{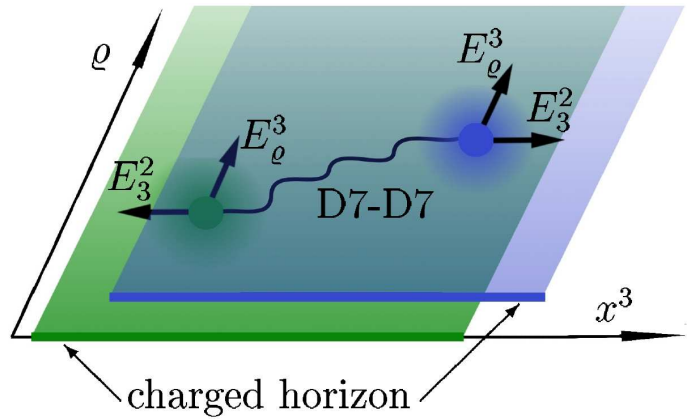}
  \caption{Sketch of our string setup: The figure shows the two
    coincident D7 branes stretched from the black hole horizon to the
    boundary as a
    green and a blue plane, respectively.  
Strings spanned from the horizon of the AdS
    black hole to the D$7$-branes  induce 
    a charge at the horizon
    \cite{Erdmenger:2008yj,Kobayashi:2006sb,Karch:2007br}. However, above a
    critical charge density, the strings charging the horizon recombine to
    D$7$-D$7$ strings. These D$7$-D$7$ strings are shown in the figure.
    Whereas the fundamental strings stretched between the horizon and the
    D$7$-brane are localized near the horizon, the D$7$-D$7$ strings propagate
    into the bulk balancing  the flavorelectric and 
    gravitational, i.e. tension forces (see text).  
    Thus these D$7$-D$7$ strings distribute
    the isospin charges along the AdS radial coordinate, leading to a
    stable configuration of reduced energy. This configuration of
    D$7$-D$7$ strings corresponds to a superconducting condensate.}
  \label{fig:stringpic}
\end{figure}

Let us first describe the unstable configuration in absence 
of the field $A^1_3$.
As known from~\cite{Karch:2007br,Kobayashi:2006sb,Erdmenger:2008yj}, the
non-zero field $A^3_0$ is induced by fundamental strings which are stretched
from the D$7$-brane to the horizon of the black hole. 
In the subsequent we call these strings `horizon strings'. 
Since the 
tension of these strings  would increase as they move to the boundary, they are
localized at the horizon, \ie the horizon is effectively charged under
the isospin charge given by (\ref{eq:7}). By increasing the horizon
string density, 
the isospin charge on the D7-brane at the horizon and therefore the energy
  of the system grows. In~\cite{Erdmenger:2008yj}, the critical 
density was found beyond which this setup becomes unstable. In this case, the
strings would prefer to move towards the boundary due to the repulsive
force on their charged endpoints generated by the flavorelectric field
$E^3_\vrho=F^3_{0\vrho}=-\del_\vrho A^3_0$. 

The setup is now stabilized by the new non-zero field $A^1_3$.
This field  is induced by D$7$-D$7$ strings moving in the $x^3$ direction.
This movement of the strings may be interpreted as a current in $x^3$
direction which induces  the magnetic field
$B^1_{3\vrho}=F^1_{3\vrho}=-\del_\vrho A^1_3$. Moreover, the non-Abelian
interaction between the D$7$-D$7$ strings and the horizon strings induces an
flavorelectric field $E_3^2=F^2_{30}=\gamma/\sqrt{\lambda}A_0^3A_3^1$.

\begin{figure}
  \centering
  \psfrag{rho}{$\rho$}
  \psfrag{A30}{$\At_0^3$}
  \psfrag{A13}{$\At_3^1$}
  \psfrag{p30}{$\pt_0^3$}
  \psfrag{p13}{$\pt_3^1$}
  \subfigure[]{\includegraphics[width=0.4\linewidth]{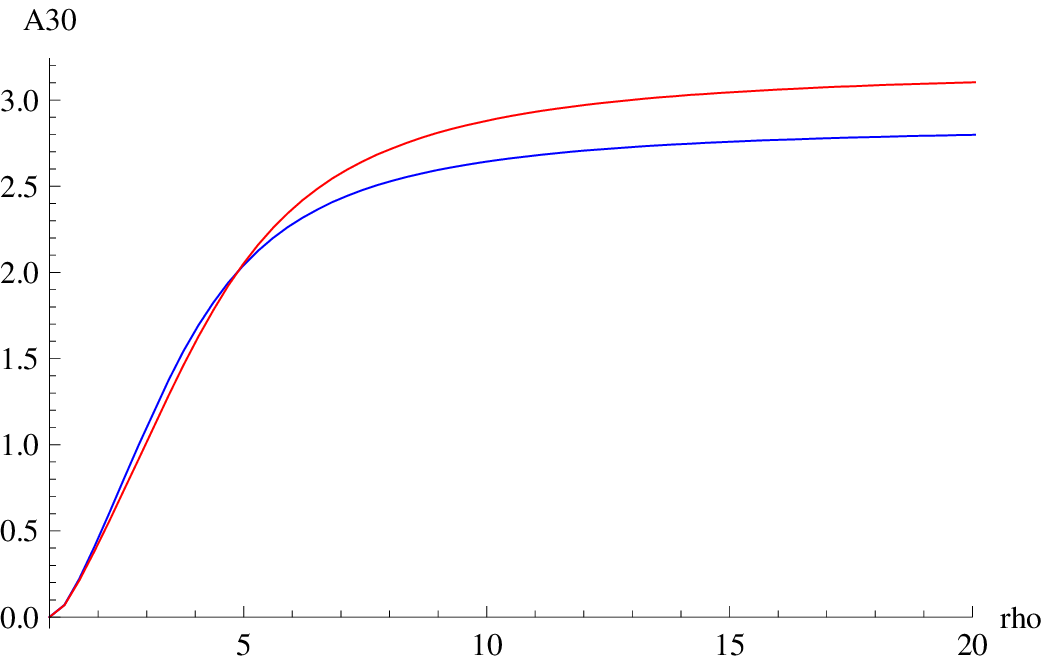}}
  \hfill
  \subfigure[]{\includegraphics[width=0.4\linewidth]{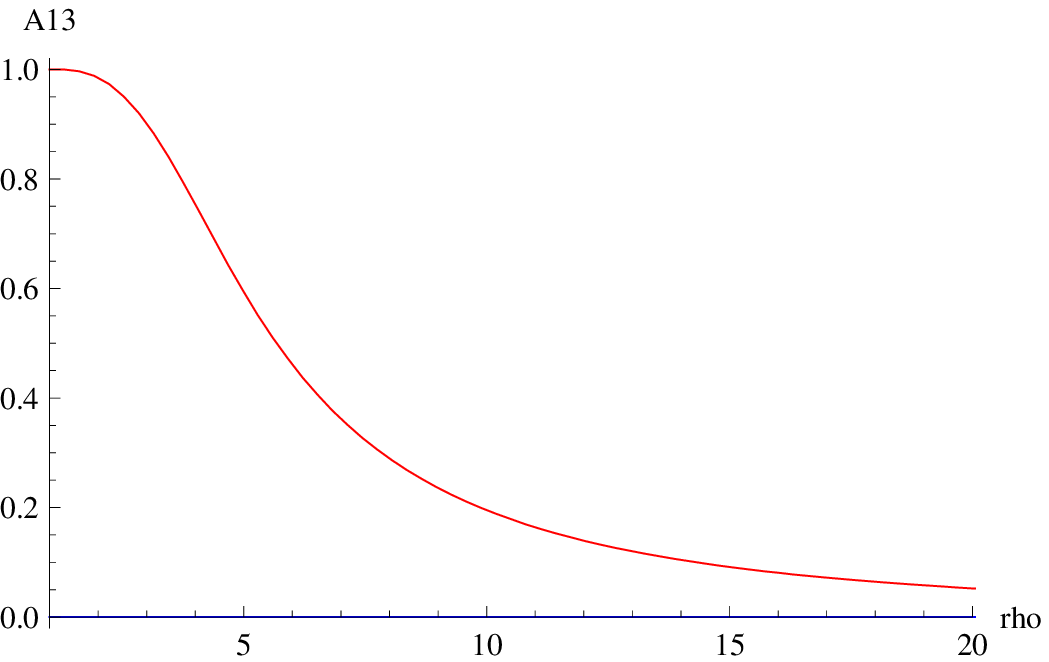}}
  \subfigure[]{\includegraphics[width=0.4\linewidth]{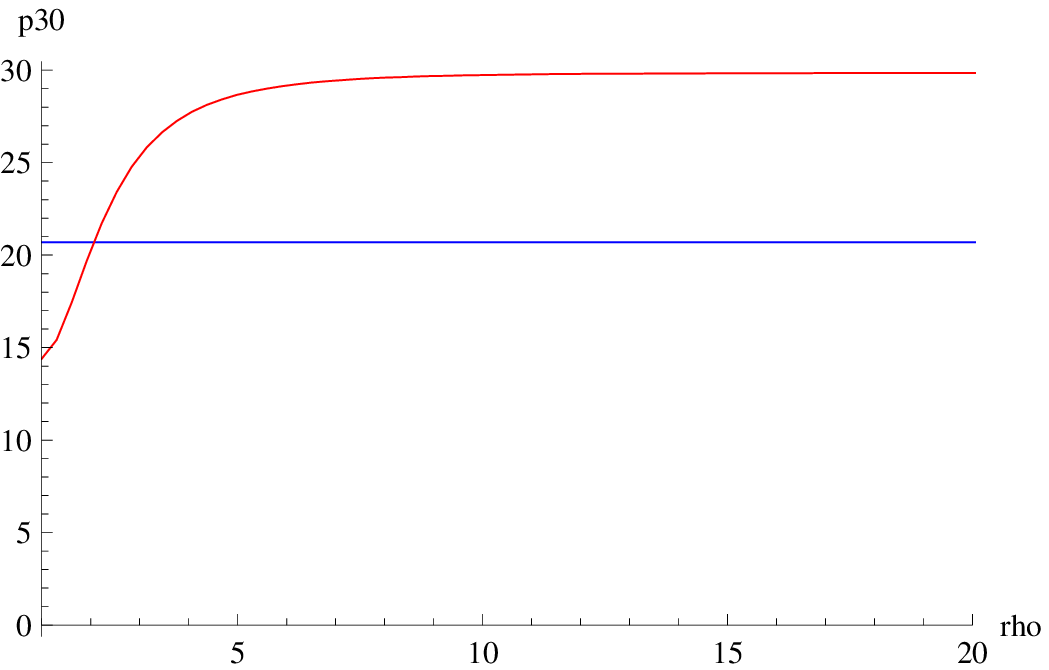}}
  \hfill
  \subfigure[]{\includegraphics[width=0.4\linewidth]{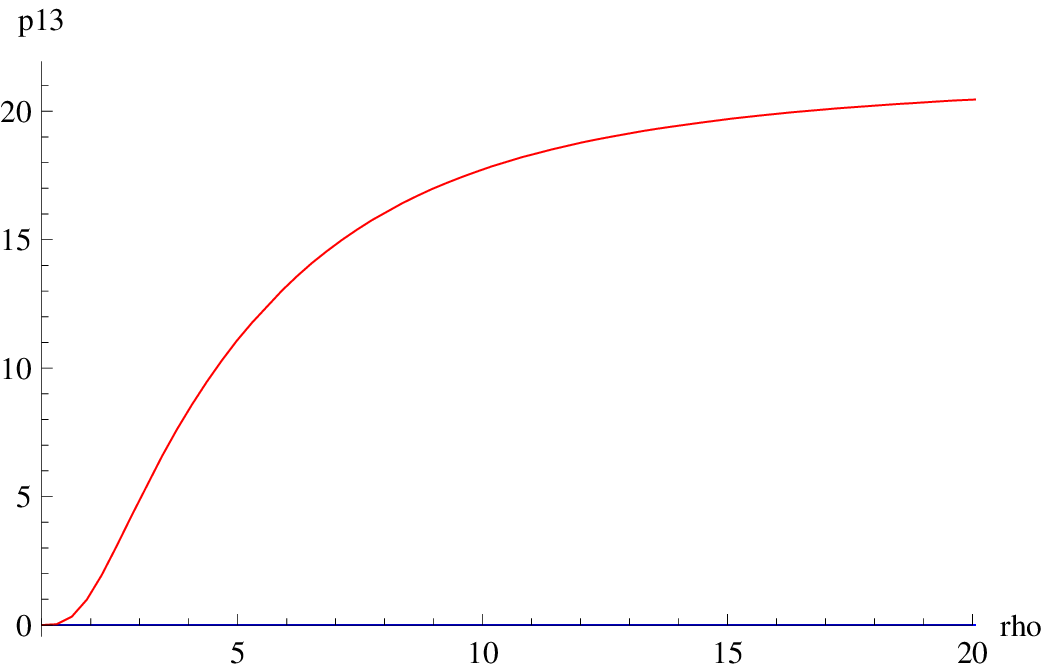}}
  \caption{Profiles of the relevant dimensionless gauge fields $\At$ on the D$7$-branes and
    their dimensionless conjugate momenta $\pt$ versus the dimensionless AdS radial coordinate
    $\rho$ near the horizon at $\rho=1$. The different curves correspond to
    the temperature $T=T_c$ (blue) and $T\approx 0.9T_c$ (red). The plots
    are obtained at zero quark mass $m=0$ and by using the adapted
    symmetrized trace prescription. Similar plots may also be obtained at
    finite mass $m\not=0$ and by using the DBI action expanded to fourth
    order in $F$. These plots show the same features: (a) The gauge field $\At_0^3$ increases monotonically
    towards the boundary. At the boundary, its value is given by the
    dimensionless chemical potential $\mut$. (b) The gauge field $\At_3^1$ is zero for $T\ge T_c$. For
    $T<T_c$, its value is non-zero at the horizon and decreases monotonically
    towards the boundary where its value has to be zero. (c) The conjugate momentum
    $\pt_0^3$ of the gauge field $\At_0^3$ is constant for $T\ge T_c$. For
    $T<T_c$, its value increases monotonically towards the boundary. Its boundary
    value is given by the dimensionless density $\dt_0^3$. (d) The conjugate
    momentum $\pt_3^1$ of the gauge field $\At_3^1$ is zero for $T\ge T_c$. For
    $T<T_c$, its value increases monotonically towards the boundary. Its boundary
    value is given by the dimensionless density $-\dt_3^1$.}
  \label{fig:profilesgaugefields}
\end{figure}

From the profile of the gauge fields and their conjugate momenta (see
figure~\ref{fig:profilesgaugefields}) we obtain the following: For $A_3^1=0$,
\ie in the normal phase 
($T\ge T_c$), the isospin density $\dt_0^3$ is exclusively generated at 
the horizon by the horizon strings. 
This can also be understood by the profile of the conjugate
momenta $p_0^3$ (see figure~\ref{fig:profilesgaugefields} (c)). We interpret
$p_0^3(\rho^*)$ as the isospin charge located between the horizon at $\rho=1$
and a fictitious boundary at $\rho=\rho^*$. 
In the normal phase, the momentum $p_0^3$ is
constant along the radial direction $\rho$ (see
figure~\ref{fig:profilesgaugefields}(c), blue curve),
and therefore the isospin density is
exclusively generated at the horizon. In the superconducting phase where
$A_3^1\not=0$, \ie $T<T_c$, the
momentum $p_0^3$ is not constant any more. Its value increases monotonically
towards the boundary and asymptotes to $\dt_0^3$ 
(see
figure~\ref{fig:profilesgaugefields}(c), red curve). 
Thus the isospin charge is also generated in the bulk
and not only at the horizon. This decreases the isospin charge 
at the horizon and stabilizes the system. 

Now we describe the string dynamics which distributes the
isospin charge into the bulk. Since the field $A_3^1$ induced by the D$7$-D$7$
strings is non-zero in the superconducting
phase, these strings must be responsible for
stabilizing this phase.  In the normal phase, there are only horizon
strings. In the superconducting phase, some of these strings
recombine to form D$7$-D$7$ strings which correspond to the non-zero gauge
field $A_3^1$ and carry isospin charge \footnote{Note that the
 D$7$-D$7$ strings are of the same order as the horizon
 strings, namely $N_f/N_c$, since they originate from the DBI action \cite{Karch:2007br}.}. 
There are two forces acting on the D$7$-D$7$ strings, the flavorelectric
force induced by the field $E_\vrho^3$ and the gravitational force between
the strings and the black hole. The flavorelectric force points to the
boundary while the gravitational force points to the horizon.  The
gravitational force is determined by the change in effective string 
tension, which contains the $\vrho$ dependent warp factor. The position of
the D$7$-D$7$ strings is determined by the equilibrium of these two forces.
Therefore the D$7$-D$7$ strings propagate from the horizon into the bulk and
distribute the isospin charge.

Since the D$7$-D$7$ strings induce the field $A_3^1$, they also generate the
density $\dt_3^1$ dual to the condensate $\langle J_3^1\rangle$, the
Cooper pairs. This density $\dt_3^1$
is proportional to the D$7$-D$7$ strings located in the bulk, 
in the same way as
the density $\dt_0^3$ counts the strings which carry isospin charge
\cite{Kobayashi:2006sb}. This suggest that we can also interpret $p_3^1(\rho^*)$ as the number
of D$7$-D$7$ strings which are located between the horizon at $\rho=1$ and the
fictitious boundary at $\rho=\rho^*$. The momentum $p_3^1$ is always zero at the
horizon and increases monotonically in the bulk (see
figure~\ref{fig:profilesgaugefields} (d)). Thus there are no D$7$-D$7$ strings
at the horizon.

The double importance of the D$7$-D$7$ strings is given by the fact that they
are both responsible for stabilizing the superconducting
phase by lowering the isospin charge
density, as well as being the dual of the Cooper pairs since they break the
$U(1)_3$ symmetry.


\section{Thermodynamics \& Phase Transition}
\label{sec:thermo}
In this section we study the thermodynamics of the fundamental matter sector
which is dual to the thermal contributions of the D$7$-branes. According the
AdS/CFT dictionary the partition function $Z$ of the boundary field theory is given
in terms of the Euclidean on-shell supergravity action $I_{\text{on-shell}}$,
\begin{equation}
  \label{eq:partitionfct}
  Z=\ee^{-I_{\text{on-shell}}}\,.
\end{equation}
Thus the thermodynamical potential, \ie in the grand canonical ensemble the grand
potential $\Omega$, is proportional to the Euclidean on-shell supergravity
action,
\begin{equation}
  \label{eq:grandpot}
  \Omega=-T\ln Z=T I_{\text{on-shell}}\,.
\end{equation}
To calculate the thermal contributions of the D$7$-branes $\Omega_7$, we have
to determine the Euclidean version of the DBI action \eqref{eq:DBI} on-shell.

\subsection{Adapted symmetrized trace prescription}
\label{sec:adapt-symm-trace-1}
Now we calculate the Euclidean on-shell action using the adapted symmetrized
trace prescription defined in section \ref{sec:dbi-action-equations}. First we perform
a Wick rotation in the time direction. Next we renormalize the action by
adding appropriate counterterms $I_{\text{ct}}$ (see \cite{Karch:2005ms} for a
review). Since the action \eqref{eq:DBIchangedstr} has the same structure as
the DBI action in \cite{Mateos:2007vn,Kobayashi:2006sb,Erdmenger:2008yj} and
since in these cases the counterterms do not depend on the finite densities,
we can write the counterterms as in
\cite{Mateos:2007vn,Kobayashi:2006sb,Erdmenger:2008yj}, 
\begin{equation}
  \label{eq:Ict}
  I_{\text{ct}}=-\frac{\caln_\lambda}{4}
  \left[\left(\rho_{\text{max}}^2-m^2\right)^2-4mc\right]\,,
\end{equation}
where $\rho_{\text{max}}$ is the UV-cutoff and
\begin{equation}
  \label{eq:Nlambda}
  \caln_\lambda=\frac{T_{D7}V_3\mathrm{vol}(S^3)N_f\vrho_H^4}{4T}=\frac{\lambda N_cM_fV_3T^3}{32}\,,
\end{equation}
with Minkowski space volume $V_3$. Then the renormalized Euclidean on-shell
action $I_R$ may simply be written as
\begin{equation}
  \label{eq:IR}
  \frac{I_R}{\caln_\lambda}=G(m,\mut)-\frac{1}{4}
  \left[\left(\rho_{\text{min}}^2-m^2\right)^2-4mc\right]\,,
\end{equation}
where $\rho_{\text{min}}$ determines the minimal value of the coordinate
$\rho$ on the D$7$-branes, \ie $\rho_{\text{min}}=1$ for black hole embeddings
which we consider exclusively in this paper, and 
\begin{equation}
  \label{eq:G}
      G(m,\mut)=\int_{\rho_{\text{min}}}^\infty\!\dd\rho\:
  \Bigg(\rho^3f\ft(1-\chi^2)\Upsilon(\rho,\chi,\At)-\left(\rho^3-\rho m\right)\Bigg)\,.
\end{equation}
In figure~\ref{fig:W7symTr} we plot the dimensionless grand potential $\calw_7$
defined as
\begin{equation}
  \label{eq:W7}
  \Omega_7=TI_R=\frac{\lambda N_fN_cV_3T^4}{32}\calw_7\,.
\end{equation}
We also calculate the specific heat $C_7$
\begin{equation}
  \label{eq:specificheat}
  C_7=-T\frac{\del^2 F_7}{\del T^2}\,,
\end{equation}
where $F_7$ is the free energy in the canonical ensemble \footnote{Note that in
\cite{Ammon:2008fc} we considered the second derivative of the grand
potential in the grand canonical ensemble. Although not simply related to the
specific heat, that quantity also shows a discontinuity indicating the second order
phase transition.}.
Using \eqref{eq:cm}, the expression for the specific heat can be rewritten as
\begin{equation}
  \label{eq:13}
  C_7=-\frac{\lambda N_fN_cT^3}{32}
  \left(12\calf_7-6m\frac{\del\calf_7}{\del m}+m^2\frac{\del^2\calf_7}{\del m^2}\right)\,,
\end{equation}
with $F_7=({\lambda N_fN_cV_3T^4}/{32})\;\calf_7$.
In figure~\ref{fig:CvsymTr} we plot the dimensionless specific heat $\calc_7$
defined as
\begin{equation}
  \label{eq:calc7}
  C_7=\frac{\lambda N_fN_cT^3}{32}\calc_7\,.
\end{equation}
Below we discuss that our results hint towards a boson liquid being present.
We also comment on the relation of our results to \cite{Karch:2008fa}.

\subsection{Expansion of the DBI action}
\label{sec:expansion-dbi-action-2}
The thermodynamical behavior in the case of the expanded DBI action may be
determined in the same way as for the adapted symmetrized trace prescription.
As mentioned above, the counterterms are needed to regularize the action and do not
depend on the gauge fields on the brane. They are needed to regularize the
divergence which occurs due to the infinite volume of the AdS
space. For vanishing gauge fields on the brane, the action calculated using the
adapted symmetrized trace prescription and the action expanded in 
field strength $F$
coincide. Thus the counterterms discussed above may also be used to
regularize the action expanded in $F$. For the renormalized Euclidean
action we obtain
\begin{equation}
  \label{eq:IRexpand}
  \frac{I_R}{\caln_\lambda}=G(m,\mut)-\frac{1}{4}
  \left[\left(\rho_{\text{min}}^2-m^2\right)^2-4mc\right]\,,
\end{equation}
where
\begin{equation}
  \label{eq:Gexp}
  G(m,\mut)=\int_{\rho_{\text{min}}}^\infty\!\dd\rho
  \Bigg[\sqrt{-G}\left(1+\frac{\calt_2}{2}-\frac{\calt_4}{8}\right)-\left(\rho^3-\rho m\right)\Bigg]\, ,
\end{equation}
with $\calt_2$ and $\calt_4$ as defined in equations (\ref{eq:DBIaxpand}) and (\ref{eq:calt}).
In figure~\ref{fig:W7dbiExp} we also plot the dimensionless grand potential $\calw_7$
obtained by the action expanded to fourth order in $F$ and defined as
\begin{equation}
  \label{eq:grandpotexp}
  \Omega_7=TI_R=\frac{\lambda N_fN_cV_3T^4}{32}\calw_7\,.
\end{equation}
Note that we have not computed the Legendre-transformed action 
for the DBI action expanded to fourth order due to technical complications discussed in 
section \ref{sec:expansion-dbi-action}. Thus we do not have direct access to the 
quantities in the canonical ensemble, such as the free energy needed 
to compute the specific heat. A calculation exploiting thermodynamical 
relations between the ensembles is postponed to future work.

\subsection{Thermodynamical consistency}
\label{sec:thermoConsistency}
In the following we confirm that both definitions of the grand potential, \eqref{eq:W7} and \eqref{eq:grandpotexp},
give consistent thermodynamics. For this purpose, we have to vary the grand
potential with respect to the thermodynamical variables. In the gravity setup
this variation is induced by a variation of the gravity fields. Here we only
consider a variation of the gauge field $A^a_\mu$, since the variation of the embedding
function $\chi$ is the same as in \cite{Kobayashi:2006sb}. Using the equations
of motion, the variation of the dimensionless grand potential with respect to
the gauge field $A_\mu^a$ reduces to a boundary term, 
\begin{equation}
  \label{eq:varygrandpot}
  \begin{split}
    \delta\calw_7&=\left[\frac{\del\call}{\del(\del_\rho \At_0^3)}\delta
    \At_0^3+\frac{\del\call}{\del(\del_\rho \At_3^1)}\delta
    \At_3^1\right]_{\rho_{\text{min}}}^\infty\\
  &=\pt_0^3(\rho_{\text{min}})\delta \At_0^3(\rho_{\text{min}})+\pt_3^1(\rho_{\text{min}})\delta \At_3^1(\rho_{\text{min}})
  -\pt_0^3(\infty)\delta \At_0^3(\infty)-\pt_3^1(\infty)\delta \At_3^1(\infty)\,.
  \end{split}
\end{equation}
Here $\call$ is given by the integrand of the grand potential. Using the
asymptotic behavior of the gauge fields
\eqref{eq:asymh},~\eqref{eq:asymb},~\eqref{eq:asymhexp}~\eqref{eq:asymb}, the
contribution of the fields $\At_3^1$ and $\pt_3^1$ vanishes and the result
coincides with \cite{Kobayashi:2006sb},
\begin{equation}
  \label{eq:varygrandpot2}
  \delta\calw_7=-\pt_0^3(\infty)\delta \At_0^3(\infty)=-\dt_0^3\delta\mut\,.
\end{equation}
Therefore we confirm that $\Omega_7$ is the thermodynamical potential in the
grand canonical ensemble, namely the grand potential. We also see that the
density $\dt_3^1$ is not a thermodynamical variable.

\subsection{Results  \& Comparison of the two prescriptions}
\label{sec:numerical-results}
We now compare the results obtained by evaluating the non-Abelian DBI action
using the adapted symmetrized trace prescription or the expansion to fourth order.

\begin{figure}
  \centering
  \psfrag{T}[t]{$\frac{T}{T_c}$}
  \psfrag{W7}[c]{$\calw_7$}  
  \includegraphics[width=0.6\linewidth]{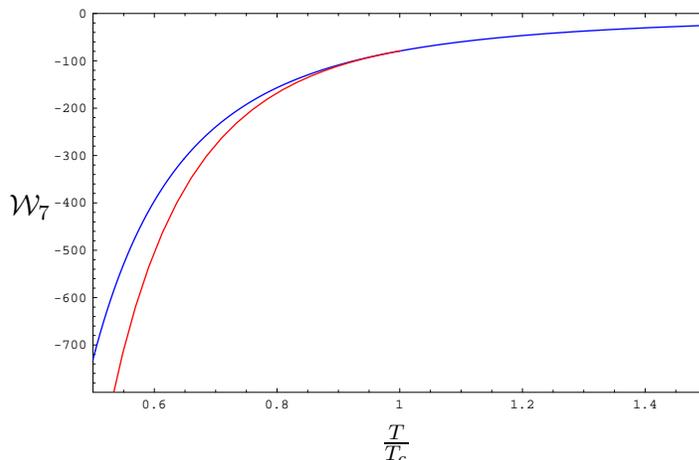} 
  \caption{The dimensionless grand canonical potential $\calw_7$
   calculated using the adapted symmetrized trace prescription versus temperature 
   at zero quark mass $M_q=0$:
   Below $T=T_c$ the superconducting phase (red line) is thermodynamically
   preferred over the normal phase (blue line).
   } 
 \label{fig:W7symTr}
\end{figure}

Figure~\ref{fig:W7symTr} shows that for non-zero gauge field $A^1_3$, a phase
transition occurs. The presence of this phase transition does not depend on the prescription
used, as seen from the comparison of the potential derived using the adapted symmetrized trace 
prescription (see figure~\ref{fig:W7symTrzoom}) with the results derived
from the expanded DBI action (see figure~\ref{fig:W7dbiExp}).
The transition is second order since there is a
discontinuous step in the specific heat $\calc_7$, see figure~\ref{fig:CvsymTr}. 

\FIGURE{
  \centering
  \psfrag{T}[t]{$\frac{T}{T_c}$}
  \psfrag{W7}[c]{$\calw_7$}  
  \includegraphics[width=0.6\linewidth]{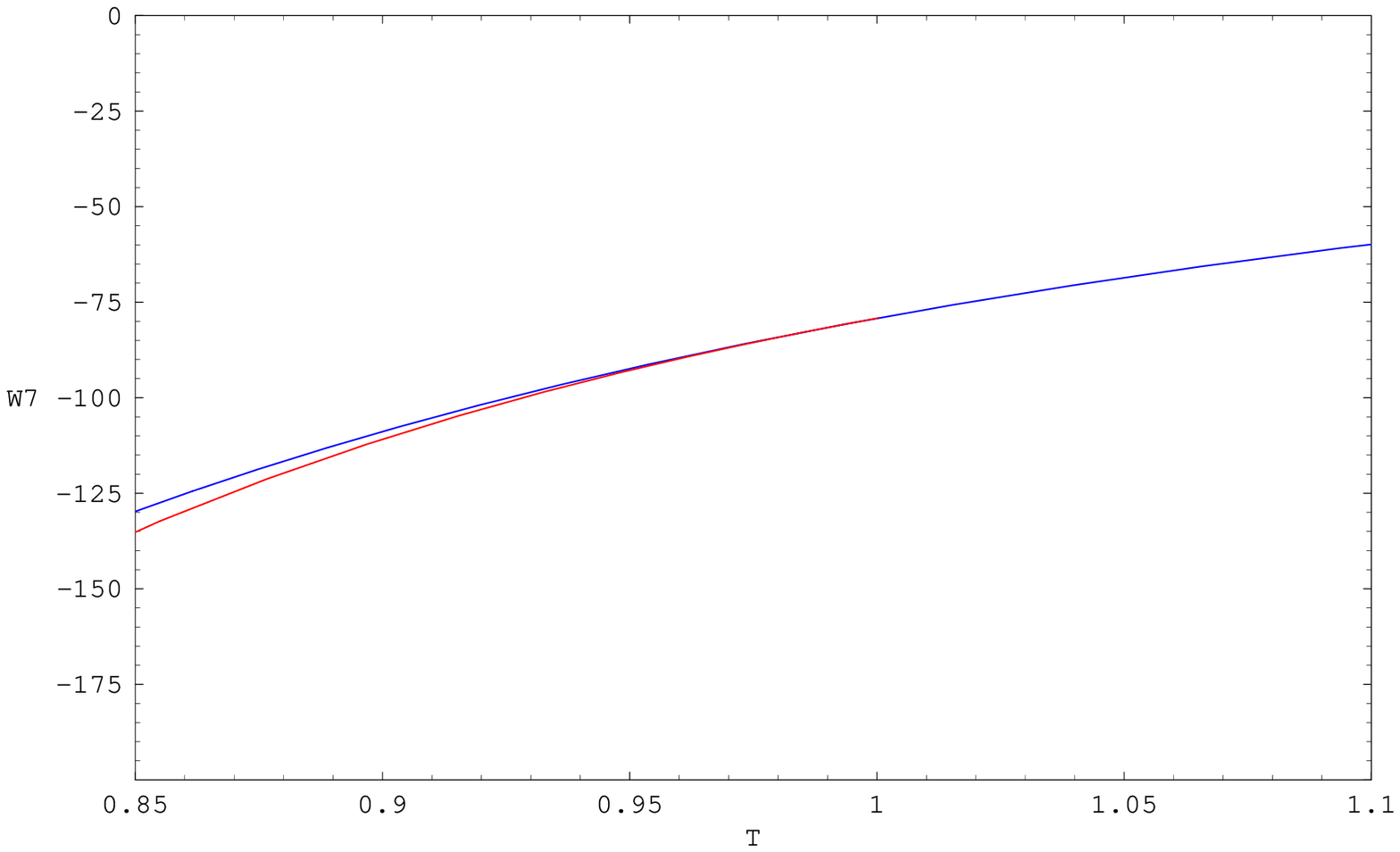}
  \caption{Same as figure~\ref{fig:W7symTr} close to the phase transition.} 
 \label{fig:W7symTrzoom}
}

\FIGURE{
  \centering
  \psfrag{T}[t]{$\frac{T}{T_c}$}
  \psfrag{W7}[c]{$\calw_7$}    
  \includegraphics[width=0.6\linewidth]{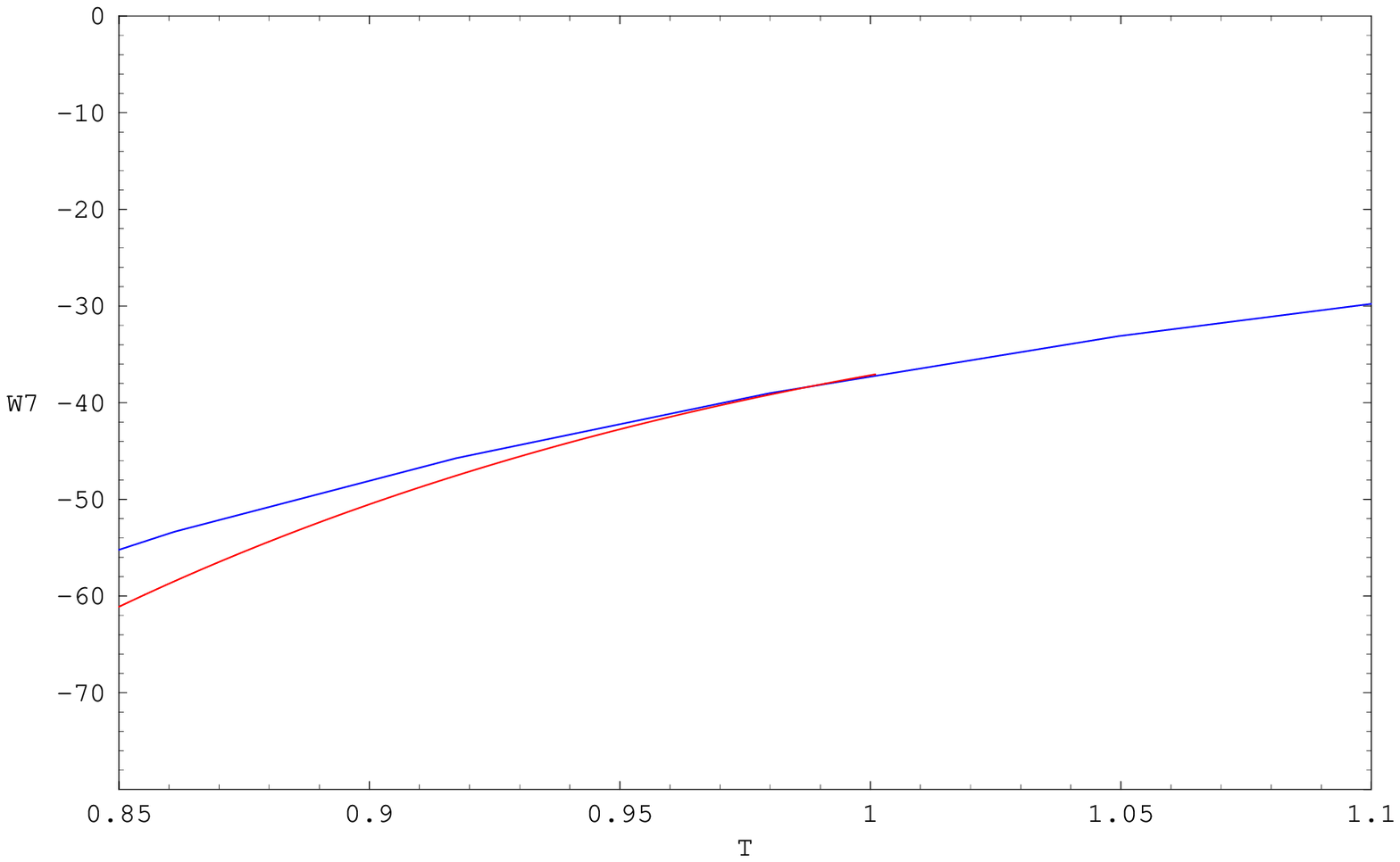}  
  \caption{Grand canonical potential computed from the expanded DBI action
  at vanishing quark mass $M_q=0$.
  The qualitative behavior is in agreement with the results from 
  the adapted symmetrized trace prescription (see figure~\ref{fig:W7symTrzoom}).
   } 
 \label{fig:W7dbiExp}
}

The temperature scale in the massless case is defined by
\begin{equation}
  \label{eq:5}
  \frac{T}{T_c}=\frac{\mut_c}{\mut}=
  \left(\frac{(\dt^3_0)_c}{\dt^3_0}\right)^{\frac{1}{3}}\,,
\end{equation}
where $\mut_c\approx 2.85$ and $(\dt^3_0)_c\approx20.7$ as obtained from the
adapted symmetrized trace prescription (see section \ref{sec:change-str-prescr}).
For the expanded DBI action we equivalently obtain 
$\mut_c\approx 2.48$ and $(\dt_0^3)_c\approx 9.03$. 
In the massive case we fix a quark mass $M_q$, then a chemical potential $\mu$
measured in units of $M_q$. Thus the temperature scale is defined as
\begin{equation}
\frac{T}{T_c}=\frac{m_c}{m}\, ,
\end{equation}
where $m_c$ is the value of the mass parameter at which the 
phase transition occurs.

\FIGURE{
  \centering
  \psfrag{T}[t]{$\frac{T}{T_c}$}
  \psfrag{mind31}{$-\dt_3^1$}  
  \includegraphics[width=0.6\linewidth]{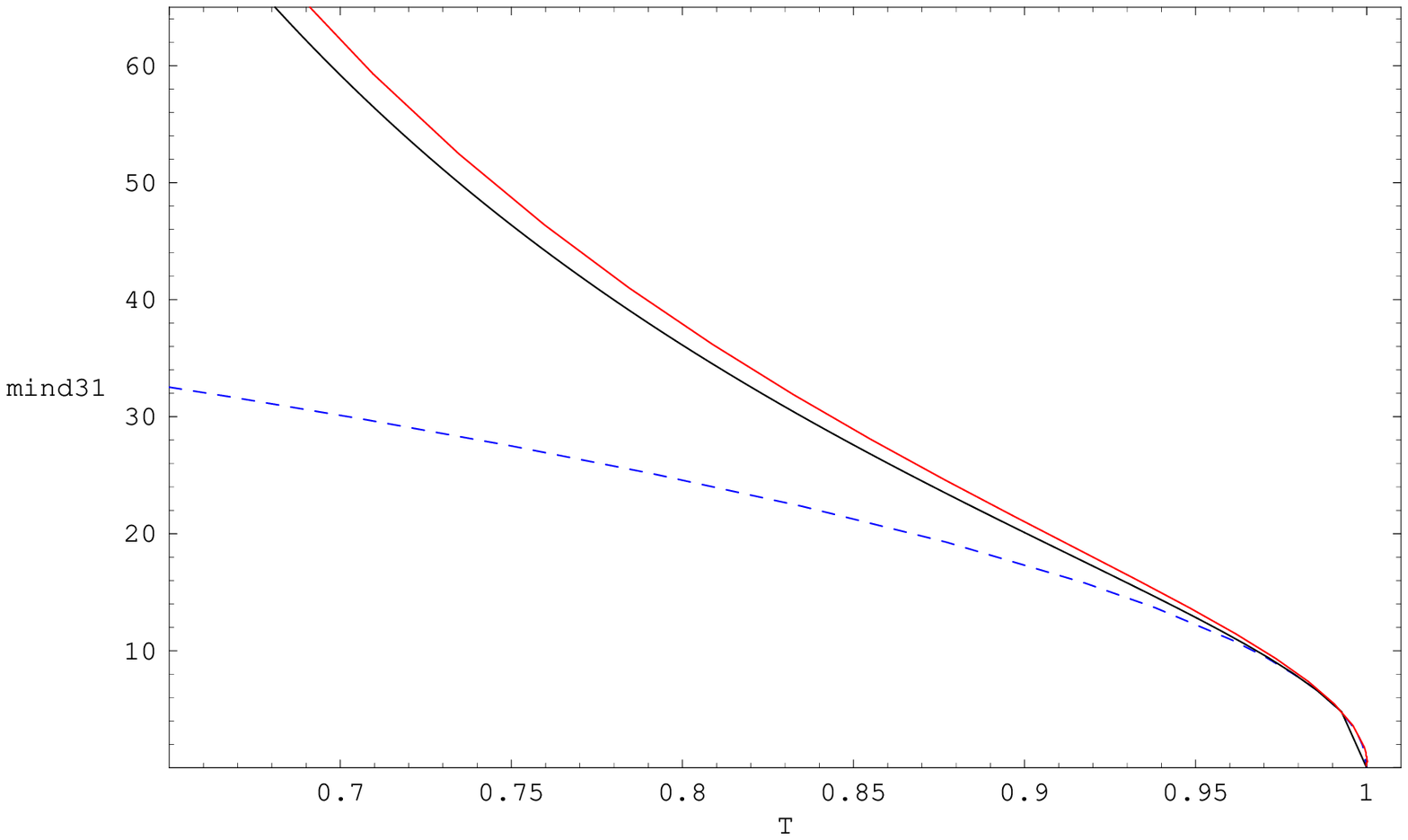}  
  \caption{The order parameter $\dt_3^1$ defined in (\ref{eq:asymb}) and obtained from the 
  adapted symmetrized trace prescription versus temperature $T$: The
  case of vanishing quark mass (red curve) shows the same behavior
  near $T_c$ as that at finite mass where $\mu/M_q=3$ is fixed. 
  In both cases the order parameter vanishes with a critical exponent of $1/2$.
  This is visualized by the fit $55 (1-T/T_c)^{1/2}$ (dashed blue curve).
   } 
 \label{fig:d13symTr}
}

From the behavior of the order parameter $\dt^1_3$, see in
figure~\ref{fig:d13symTr}, near the critical temperature, we
obtain the critical exponent of the transition to be $1/2$, which coincides
with the result obtained from Landau theory. It is also determined
numerically from a fit within 10 percent accuracy, see blue dashed curve 
in figure~\ref{fig:d13symTr}. This exponent is independent from the
quark mass as seen from the black curve at $\mu/M_q=3$ shown in
figure~\ref{fig:d13symTr}. We have also confirmed numerically that the
critical exponent of the order parameter $\dt_3^1$ does not depend 
on the prescription (adapted symmetrized trace or expanded DBI). 
Since the order parameter
$\dt^1_3$ and the density $\dt^3_0$ increase rapidly as $T\to 0$, we expect
that the probe approximation breaks down near $T=0$. 

\begin{figure}
  \centering
  \psfrag{T}[t]{$\frac{T}{T_c}$}
  \psfrag{ds}{$\dt_s$}    
  \includegraphics[width=0.6\linewidth]{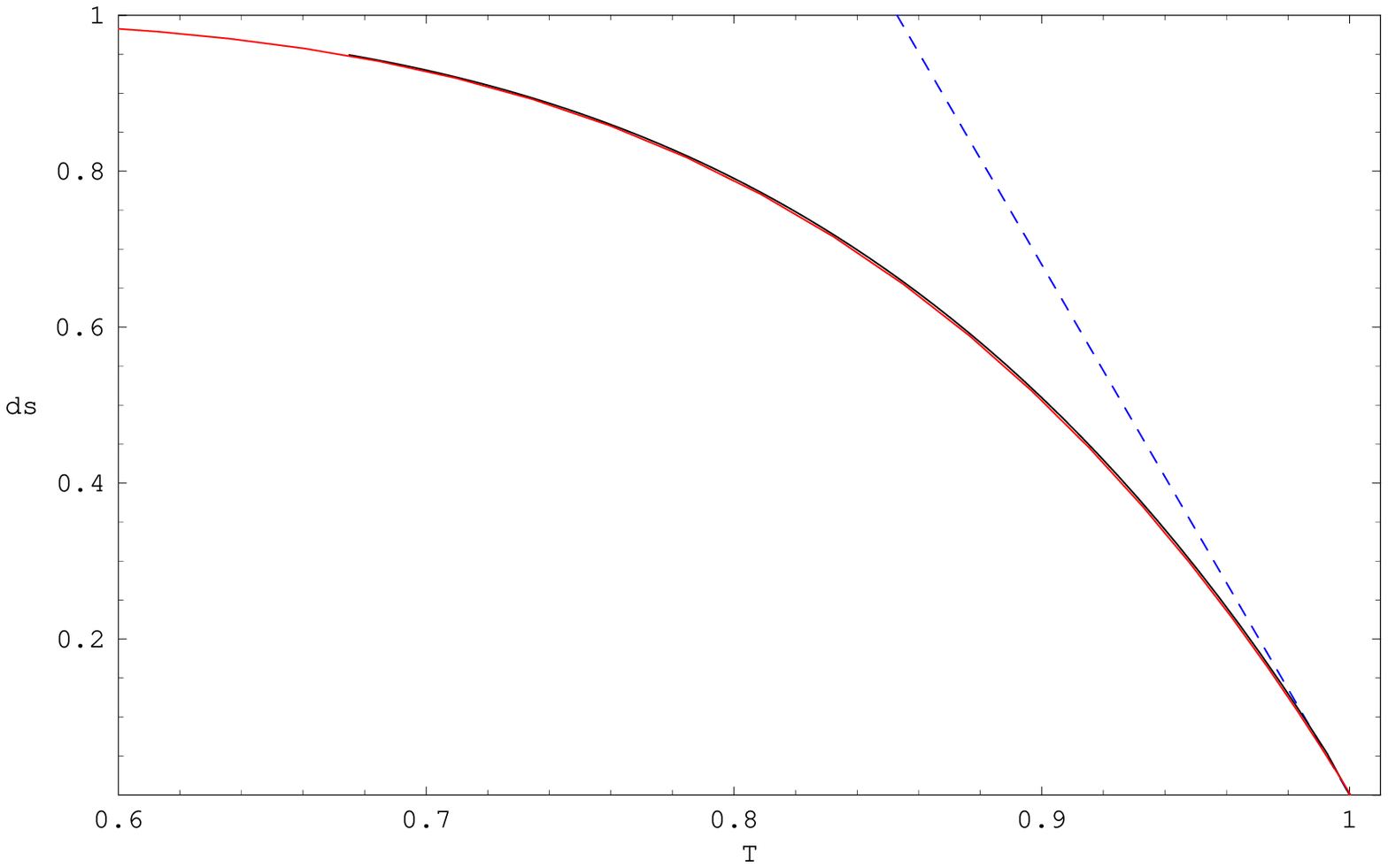}  
  \caption{Superconducting density $\dt_s=(\dt_0^3-c_0)/\dt_0^3$ versus temperature $T$:
  In both, the massless (red curve) and the massive case at $\mu/M_q=3$ (blue curve), the superconducting density $\dt_s$
  vanishes linearly at the critical temperature. This is visualized by the 
  fit $6.8 (1-T/T_c)$ (dashed blue curve).
   } 
 \label{fig:dssymTr}
\end{figure}

We define the density of superconducting charges $\dt_s$ in the following way
\begin{equation}\label{eq:defds}
\dt_s=\frac{\pt_0^3(\rho_B)-\pt_0^3(\rho_H)}{\pt_0^3(\rho_B)}=\frac{\dt_0^3-c_0}{\dt_0^3}\, .
\end{equation}
This identification was suggested in~\cite{Gubser:2008wv}. Beyond the 
arguments given there, we have the advantage of exactly knowing the dual 
field theory interpretation and having a string picture, see section
\ref{sec:string-theory-pict}. The numerator of equation \eqref{eq:defds}
counts the isospin charges present in the bulk excluding those localized at the black hole horizon 
$p_0^3(\rho_H)=c_0$. Following the ideas described in section 
\ref{sec:string-theory-pict}, these charges may be identified with the 
D$7$-D$7$ strings in the bulk, which are dual to the Cooper pairs.
As expected from Ginzburg-Landau theory, this superconducting 
density $\dt_s$ vanishes linearly at $T_c$ in massive and massless cases
regardless of the prescription chosen, see figure~\ref{fig:dssymTr}.

\begin{figure}
  \centering
  \psfrag{T}[t]{$\frac{T}{T_c}$}
  \psfrag{C}[r]{$\calc_7$}    
  \includegraphics[width=0.6\linewidth]{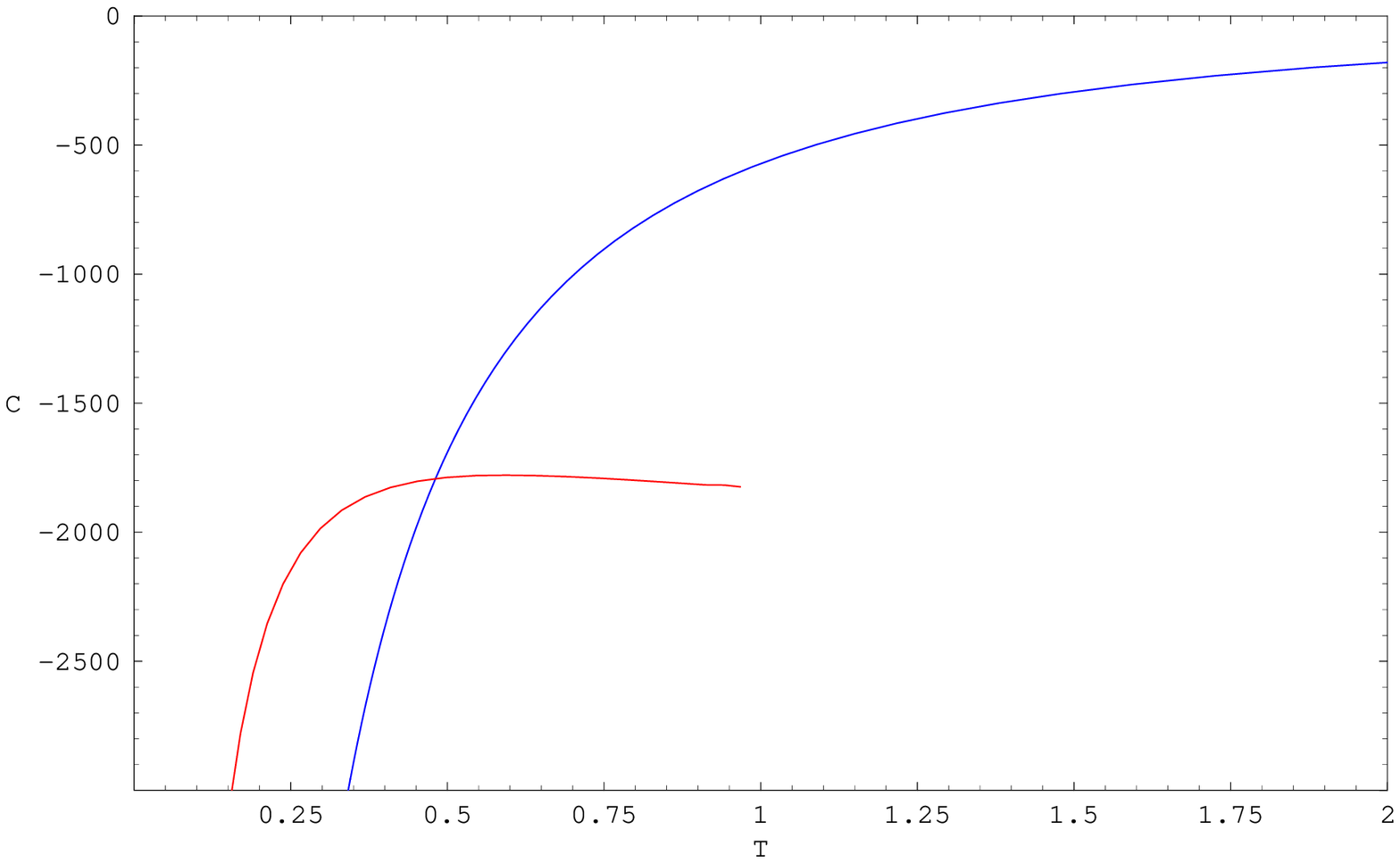}  
  \caption{The flavor brane contribution to the specific heat 
  as computed from the adapted symmetrized trace prescription in the massless
  case. The blue line corresponds to the normal phase with $A_3^1=0,$ while the
  red one corresponds to the superconducting phase with $A_3^1\not=0$.
  Note that the total specific heat is always positive although the
  flavor brane contribution is negative. The divergences near $T=0$ in both
  phases can be attributed to the missing backreaction in our setup. We read of 
  from the  numerical result that near the critical temperature, the dimensionless 
  specific heat is constant in the superconducting phase. This implies that the
  dimensionful specific heat is proportional to $T^3$. This temperature
  dependence is characteristic for Bose liquids.
   } 
 \label{fig:CvsymTr}
\end{figure}

In figure~\ref{fig:CvsymTr} the dimensionless 
specific heat computed from the adapted
symmetrized trace prescription for massless quarks is shown 
versus the temperature. It clearly shows a discontinuity at the phase transition, 
stressing that the observed transition is second order. 
Note that the dimensionful specific heat is given by
$C_7\propto T^3 \calc_7$, where $\calc_7$ is the dimensionless
specific heat shown in figure~\ref{fig:CvsymTr}.

The blue curve in figure \ref{fig:CvsymTr} corresponds to the normal phase while the red one 
shows the behavior of the superconducting phase. A general feature 
of superfluids, including charged superfluids as in the present case, is the
behavior of the specific heat near $T=0$. For Bose liquids the heat capacity
has cubic temperature dependence. In Fermi liquids, a linear temperature 
dependence dominates near $T=0$. In our numerical results 
(figure~\ref{fig:CvsymTr}) we see a dramatic change of specific heat in the normal phase (blue curve) 
in sharp contrast to the constant behavior, corresponding to $C_7\propto T^3$, of the superconducting
phase (red curve) close to the critical temperature. The divergent behavior
close to $T=0$ is due to the probe approximation which we use. Although our numerics do
not allow to extend the calculation up to $T=0,$ it is tempting to extrapolate
the observed behavior towards $T=0$ and to conclude that our setup in the
superconducting phase is a Bose liquid. 

Recently a new type of quantum liquid has been discussed in 
\cite{Karch:2008fa}, where the authors
consider a brane setup at zero mass in the limit of infinite baryon density, \ie 
at zero temperature. In that case the heat capacity was found to be proportional
to $T^6$, neither typical for a Fermi nor for a Bose liquid.   

In the following we argue that our result found above suggests to consider the
possibility of a phase transition also in the baryonic setup
\cite{Karch:2008fa}. The baryonic setup of \cite{Karch:2008fa} may be related to
our normal phase by a symmetry between pure baryonic and pure isospin
configurations found in \cite{Erdmenger:2008yj}. Thus the specific heat in the baryonic
setup and in our normal-conducting phase coincide. Since in our isospin case there is a
phase transition to the superconducting phase which shows the usual Bose
liquid behavior, we suggest the possibility of a similar phase transition
in the baryonic setup of \cite{Karch:2008fa}. The
possible new phase may lead to the usual behavior of a Bose or Fermi liquid.

\FIGURE{
  \centering
  \includegraphics[width=0.6\linewidth]{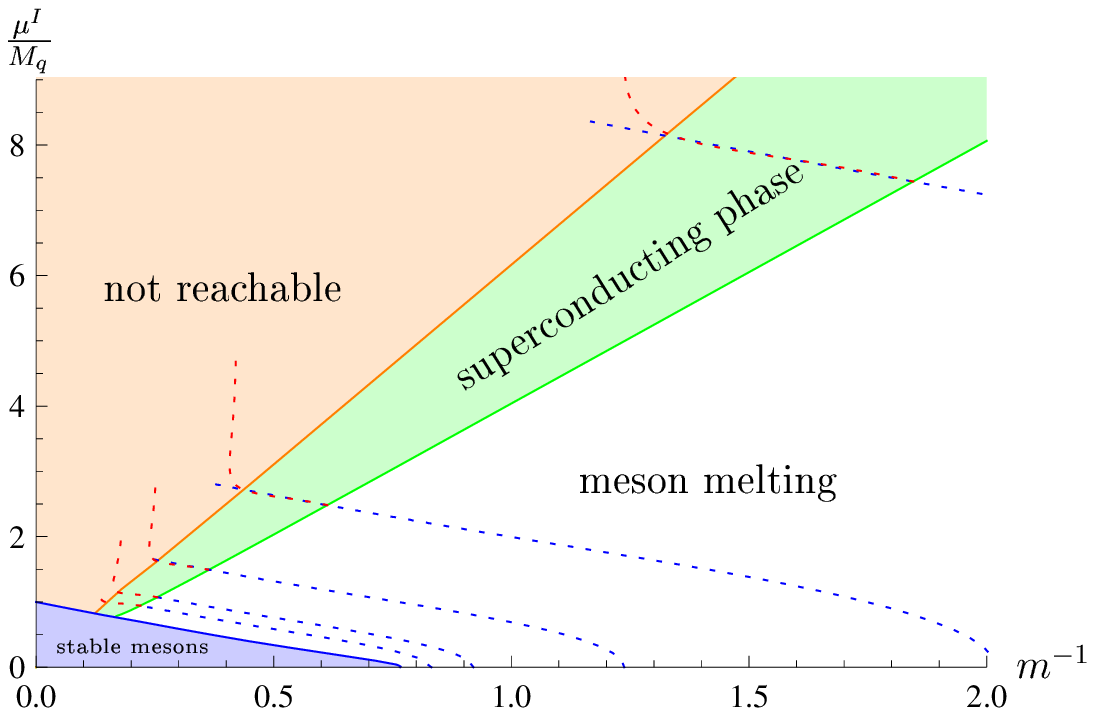}
  \caption{Phase diagram for fundamental matter with mass $m= 2 M_q / (\sqrt{\lambda} T)$ obtained with the
    adapted symmetrized trace prescription: The blue, white and green regions
    are the same as in figure~\ref{fig:phasediagram}, but with the
    unstable normal 
    phase replaced with the superconducting phase. The dotted curves
    correspond to lines at finite mass, \ie constant $\chi_0$. These curves
    are parametrized by the density $\dt_0^3$. Along the blue curves the field
    $A_3^1$ is zero while along the red ones the field $A_3^1$ is non-zero.
    The endpoints of the red curves determine the second order phase transition to the
    superconducting phase. The dotted, red curves diverge inside the
    superconducting phase since the backreaction of the condensate on the background
    is not considered. This divergence determines the boundary of the orange
    region which is not reachable without backreaction.}
  \label{fig:phasediagramfull}
}

Summarizing our thermodynamical results in a phase diagram, we obtain
figure~\ref{fig:phasediagramfull}. The choice of the calculational method --
adapted symmetrized trace or expansion of the DBI to fourth order -- 
does not change the qualitative structure of the phase diagram. 
The blue phase indicates the known region of stable mesons
surviving the deconfinement transition. It is separated 
from the white meson melting region by the meson melting transition (blue line),
see
\cite{Faulkner:2008hm,Babington:2003vm,Mateos:2007vn,Kobayashi:2006sb,Mateos:2007vc,Karch:2007br,Hoyos:2006gb,Erdmenger:2008yj}.
Above a critical isospin density marked by the green line, a
flavor-superconducting phase forms. At even higher isospin density, our
approach, which does not include the back-reaction of the D$7$-brane, gives
diverging order parameters, signalling the breakdown of this approach. This
particular region is indicated by orange color. Note that this behavior is
well-known to occur in systems without back-reaction. In our case, 
the gauge fields on the D$7$-brane grow arbitrarily large close to zero
temperature. This behavior will be cured by including the back-reaction, \ie
the contribution of the gauge fields to the total energy-momentum tensor. For
the Abelian Higgs model, it is shown in \cite{Hartnoll:2008kx} that similar
divergences are removed when these terms are included.


\section{Fluctuations}
\label{sec:fluctuations}
The full gauge field~$\hat A$ on the branes consists of the field~$A$ 
and fluctuations~$a$, 
\begin{equation} 
\hat A = A_0^3 \tau^3 \dd t +A_3^1 \tau^1 \dd x_3 +
  a_\mu^a \tau^a \dd x^\mu \,,
\end{equation}
where $\tau^a$ are the $SU(2)$ generators. The linearized equations of motion
for the fluctuations $a$ are obtained by expanding the DBI action in $a$ to
second order. We will analyze the fluctuations $a^3_2$ and
$X=a^1_2+\ii a^2_2$, $Y=a^1_2-\ii a^2_2$.

Including these fluctuations, the DBI action reads 
\begin{equation}
\label{eq:dbiFluctuations}  
S=-T_7\int \dd^8\xi\: \Str\sqrt{\det{[G+(2\pi\alpha')\hat F]}} \,  , 
\end{equation}    
with the non-Abelian field strength tensor 
\begin{equation}
\hat F_{\mu\nu}^a = F_{\mu\nu}^a+\check F_{\mu\nu}^a \, ,  
\end{equation} 
where the background is collected in 
\begin{equation}  
F_{\mu\nu}^a = 2\partial_{[\mu} A_{\nu]}^a 
  +\frac{\gamma}{\sqrt{\lambda}} f^{abc} A_\mu^bA_\nu^c\, ,    
\end{equation}  
and all terms containing fluctuations in the gauge field are
summed in  
\begin{equation}  
\check F_{\mu\nu}^a = 2\partial_{[\mu} a_{\nu]}^a + 
   \frac{\gamma}{\sqrt{\lambda}} f^{abc} a_\mu^b a_\nu^c + 
   \frac{\gamma}{\sqrt{\lambda}} f^{abc} (A_\mu^b a_\nu^c+a_\mu^b A_\nu^c) \, .  
\end{equation}  
Index anti-symmetrization is always defined with a factor of two
in the following way $\partial_{[\mu}A_{\nu]}= (\partial_\mu A_\nu-\partial_\nu A_\mu)/2$.

\subsection{Adapted symmetrized trace prescription}
\label{sec:adopt-symm-trace}
In this section we use the adapted symmetrized trace prescription to determine
the fluctuations about the background we discussed in section
\ref{sec:change-str-prescr}. To obtain the linearized equations of motion for
the fluctuations $a$, we expand the action~\eqref{eq:dbiFluctuations} to
second order in fluctuations, 
\begin{equation}
  \begin{split}
    S^{(2)}=-T_7\int \dd^8\xi\: \Str\Bigg[&\sqrt{-\G}
      +\frac{(2\pi\alpha')}{2}\sqrt{-\G}\G^{\mu\nu}\Fc_{\nu\mu}-\frac{(2\pi\alpha')^2}{4}
      \sqrt{-\G}
      \G^{\mu\mu'}\Fc_{\mu'\nu}\G^{\nu\nu'}\Fc_{\nu'\mu}\\
      &+\frac{(2\pi\alpha')^2}{8}
      \sqrt{-\G}\left(\G^{\mu\nu}\Fc_{\nu\mu}\right)^2\Bigg] \,. 
\end{split}
\end{equation}
As in the previous paper \cite{Erdmenger:2007ja}, we collect the metric and
gauge field background in the tensor $\G=G+(2\pi\alpha') F$. Using the
Euler-Lagrange equation, we get the linearized equation of motion for fluctuations~$a_\kappa^d$ in the form
\begin{equation}
  \label{eq:eomfluc}
  \begin{split}
    0=&\del_\lambda\Str
    \Big[\sqrt{-\G}\tau^d\Big\{\G^{[\kappa\lambda]}
    +(2\pi\alpha')\Big(\G^{\mu[\kappa}\G^{\lambda]\nu}+\frac{1}{2}\G^{\mu\nu}\G^{[\kappa\lambda]}\Big)\check
        F_{\nu\mu}\Big\}\Big]-\Str\Big[\frac{c}{\sqrt{\lambda}}f^{abd}\tau^a\sqrt{-\G}\\
        &\times\Big\{\G^{[\kappa\lambda]}(a+A)^b_\lambda+(2\pi\alpha')
          \Big(\G^{\mu[\kappa}\G^{\lambda]\nu}+\frac{1}{2}\G^{\mu\nu}\G^{[\kappa\lambda]}\Big)\check F_{\nu\mu}A^b_\lambda\Big\}\Big]\,.
  \end{split}
\end{equation}
Note that the linearized version of the fluctuation field strength used in equation~\eqref{eq:eomfluc} is given by 
\begin{equation}
  \check F_{\mu\nu}^a = 2\partial_{[\mu} a_{\nu]}^a +
  \frac{\gamma}{\sqrt{\lambda}} f^{abc} (A_\mu^b a_\nu^c+a_\mu^b A_\nu^c)
  +{\cal O} (a^2) \, .
\end{equation}
In our specific case the background tensor in its covariant form is given by
\begin{equation} 
\G_{\mu\nu}=G_{\mu\nu} \tau^0 + (2\pi\alpha')\Big ( 
 2\partial_\vrho A_0^3\delta_{4[\mu}\delta_{\nu]0}\tau^3+ 2\partial_\vrho A_3^1\delta_{4[\mu}\delta_{\nu]3}\tau^1
 + 2\frac{\gamma}{\sqrt{\lambda}}A_0^3A_3^1\delta_{0[\mu}\delta_{\nu]3}\tau^2
\Big ) \,. 
\end{equation}   
Inversion yields the contravariant form needed to compute the 
explicit equations of motion. The inverse of $\G$ is defined as
$\G^{\mu\nu}\G_{\nu\mu'}=\delta^\mu_{\mu'}\tau^0$ \footnote{We calculate the
  inverse of $\G$ by ignoring the commutation relation of the $\tau$'s because of symmetrized trace.
It is important that $\tau^a\tau^b$ must not be simplified to
$\epsilon^{abc}\tau^c$ since symmetrisation is not the same.}. The non-zero
components of $\G^{\mu\nu}$ may be found in the appendix
\ref{sec:adapt-symm-trace}.

\paragraph{Fluctuations in $a^3_2$:}
For the fluctuation $a_2^3$ with zero spatial momentum, we obtain the equation
of motion
\begin{equation}
  \label{eq:eoma32}
  (a^3_2)''+\frac{\del_\rho H}{H}(a^3_2)'-
  \Bigg[\frac{4\vrho_H^4}{R^4}\left(\frac{\G^{33}}{\G^{44}}(\m_3^1)^2+\frac{\G^{00}}{\G^{44}}\w^2\right)
  -16\frac{\del_\rho\left(\frac{H}{\rho^4f^2}\At^3_0(\del_\rho\At^3_0)(\m^1_3)^2\right)}{H\left(1-\frac{2c^2}{\pi^2\rho^4f^2}(\At^1_3\At^3_0)^2\right)}\Bigg]a^3_2=0\,,
\end{equation}
with
\begin{equation}
    \m^1_3=\frac{\gamma}{2\sqrt{2}\pi}\At^1_3\,,\qquad  H=\sqrt{\G}G^{22}\G^{44}\,.
\end{equation}

\paragraph{Fluctuations in $X=a^1_2+\ii a^2_2$, $Y=a^1_2-\ii a^2_2$:}
For the fluctuations $X$ and $Y$ with zero spatial momentum, we obtain the
coupled equations of motion
\begin{align}
  \label{eq:eomXY}
  \begin{split}
    0=&X''+\frac{\del_\rho H}{H}X'-\frac{4\vrho_H^4}{R^4}
    \Bigg[\frac{\G^{00}}{\G^{44}}\left(\w-\m^3_0\right)^2+\frac{\G^{\{03\}}}{\G^{44}}\m^1_3\w\Bigg]X+\frac{4\vrho_H^4}{R^4}
    \Bigg[\frac{\G^{\{03\}}}{\G^{44}}\m^1_3\m^3_0\\
    &+\frac{R^2}{4\vrho_H^2}\frac{\del_\rho\left[\sqrt{-\G}G^{22}\G^{\{34\}}\m^1_3\right]}{H}-\frac{\G^{33}}{2\G^{44}}\left(\m^1_3\right)^2\Bigg](X-Y)+\frac{4\vrho_H^2}{R^2}\frac{\G^{\{04\}}}{\G^{44}}\w 
    Y'\\
    &+\frac{2\vrho^2_H}{R^2}\frac{\del_\rho
      \left[\sqrt{-\G}G^{22}\G^{\{04\}}\left(\w+\m^3_0\right)\right]}{H}Y \, ,
  \end{split}\\
  \begin{split}
    0=&Y''+\frac{\del_\rho H}{H}Y'+\frac{4\vrho_H^4}{R^4}
    \left[-\frac{\G^{00}}{\G^{44}}\left(\w+\m^3_0\right)^2+\frac{\G^{\{03\}}}{\G^{44}}\m^1_3\w\right]Y-\frac{4\vrho_H^4}{R^4}
    \Bigg[
    \frac{\G^{\{03\}}}{\G^{44}}\m^1_3\m^3_0\\
    &+\frac{R^2}{4\vrho_H^2}\frac{\del_\rho\left[\sqrt{-\G}G^{22}\G^{\{34\}}\m^1_3\right]}{H}-\frac{\G^{33}}{2\G^{44}}\left(\m^1_3\right)^2\Bigg](X-Y)-\frac{4\vrho_H^2}{R^2}\frac{\G^{\{04\}}}{\G^{44}}\w
    X'\\ 
    &-\frac{2\vrho^2_H}{R^2}\frac{\del_\rho\left[\sqrt{-\G}G^{22}\G^{\{04\}}\left(\w-\m^3_0\right)\right]}{H}X \, ,
  \end{split}
\end{align}
where the component of the inverse background tensor may be found in the
appendix \ref{sec:adapt-symm-trace}, index symmetrization is defined
$\G^{\{ij\}}=(\G^{ij}+\G^{ji})/2$ and
\begin{equation}
  \m^3_0=\frac{\gamma}{2\sqrt{2}\pi}\At^3_0\,.
\end{equation}

\subsection{Expansion of the DBI action}
\label{sec:expansion-dbi-action-1}
In this section we determine the equation of motion for the fluctuation $a_2^3$
in the background determined by the DBI action expanded to fourth order in
$F$ (see section \ref{sec:expansion-dbi-action}). To obtain the
quadratic action in the field $a_2^3$, we first have to expand the DBI action
\eqref{eq:dbiFluctuations} to fourth order
in the full gauge field strength $\hat F$, and expand the result
to second order in $a_2^3$. Due to the symmetries of our setup, the equation
of motion for the fluctuation $a_2^3$ at zero spatial momentum decouples from
the other equations of motion, such that we can write down an effective
Lagrangian for the fluctuation $a_3^2$. This effective Lagrangian is given in
the appendix \ref{sec:expas-fourth-order}. The equation of motion for $a_2^3$
with zero spatial momentum determined by the Euler-Lagrange equation is
given by 
\begin{equation}
  \label{eq:eoma32exp}
  \begin{split}
  0=&(a_2^3)''+\frac{\del_\rho \H}{\H}(a_2^3)'-
  \frac{\vrho_H^4}{R^4}\Bigg[4
    \Bigg(\frac{\H^{00}}{\H^{44}}\w^2+\frac{\H^{33}}{\H^{44}}(\m_3^1)^2\Bigg)\\
    &+\frac{8}{3}\frac{\del_\rho[\sqrt{-G}G^{00}G^{22}G^{33}G^{44}\At_0^3(\del_\rho
      \At_0^3)(\m_3^1)^2]}{\H}\Bigg]a_2^3\,,
  \end{split}
\end{equation}
where
\begin{equation}
    \H=\sqrt{-G}G^{22}\H^{44}\,.
\end{equation}
We introduce the factors $\H^{ij}$ which may be found in the appendix
\ref{sec:expas-fourth-order} to emphasize the similarity to the equation of
motion obtained by the adapted symmetrized trace prescription
\eqref{eq:eoma32}. 

\subsection{Result \& Comparison of the two prescriptions}
\label{sec:result--comparison}

\subsubsection{Conductivity}
\label{sec:conductivity}
We calculate the frequency-dependent conductivity $\sigma(\omega)$ using the Kubo formula,
\begin{equation}
  \label{eq:defsigma}
  \sigma(\omega)=\frac{\ii}{\omega}G^R(\omega,q=0)\,,
\end{equation}
where $G^R$ is the retarded Green function of the current $J^3_2$ dual to
the fluctuation $a^3_2$, which we calculate using the method obtained
in~\cite{Son:2002sd}. The current $J^3_2$ is the analog to the electric
current since it is charged under the $U(1)_3$ symmetry. In real space it is
transverse to the condensate. Since this fluctuation is the only one which
transforms as a vector under the $SO(2)$ rotational symmetry, it decouples
from the other fluctuations of the system.

\begin{figure}
  \centering
  \psfrag{Resig}{$\re \sigma$}
  \psfrag{w}{$\w$}
  \includegraphics[width=0.6\linewidth]{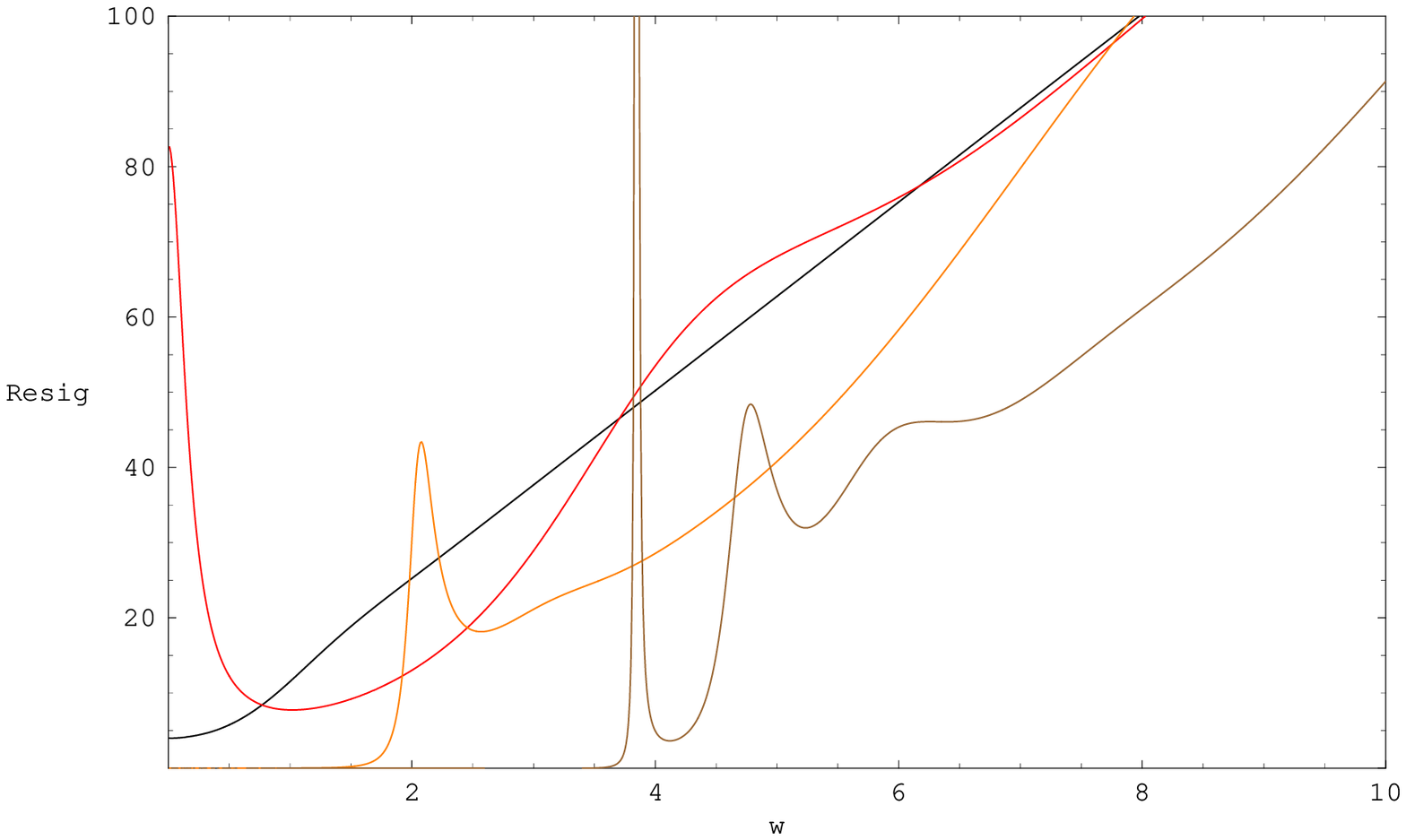}    
  \caption{Real part of conductivity, $\re \sigma,$ in units of
  $N_f N_c T/(16\pi)$ versus the dimensionless 
  frequency $\w=\omega/(2\pi T)$ for massless quarks 
  computed from the adapted symmetrized trace prescription.
  Distinct curves correspond to $T/T_c=\infty$ (black), $1$(red), 
  $0.5$ (orange) and $0.28$ (brown).  By decreasing the temperature below the
  critical one, a gap where the conductivity is approximately zero appears
  which is a characteristic feature of a superconductor. In addition prominent peaks arise.
   } 
 \label{fig:reSigsymTr}
\end{figure}

\begin{figure}
  \centering
  \psfrag{ResigS}{$\frac{\re \sigma}{4\pi\w}$}
  \psfrag{w}{$\w$}
  \includegraphics[width=0.6\linewidth]{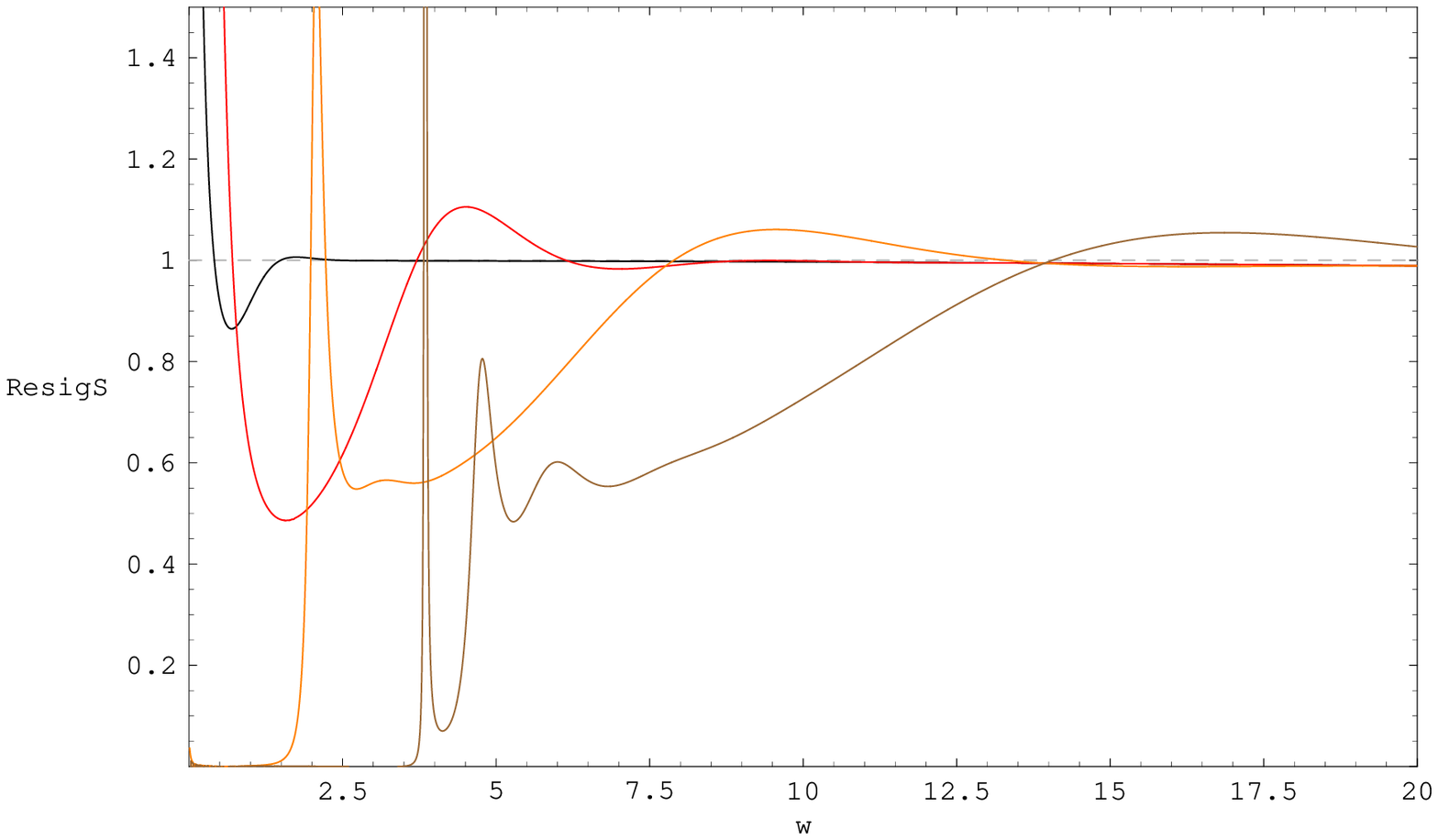}  
  \caption{Scaled real part of conductivity $\re \sigma/(4\pi \w)$ in units of
  $N_f N_c T/(16\pi)$ versus the dimensionless 
  frequency $\w=\omega/(2\pi T)$ for massive quarks at fixed $\mu/M_q=3$
  computed from the adapted symmetrized trace prescription.
  Distinct curves correspond to $T/T_c=10$ (black), $1$(red), 
  $0.5$ (orange) and $0.28$ (brown). This figure has been
  scaled to asymptote to a constant in order to show similarity
  to the lower dimensional cases computed from $AdS_4$ and 
  to show common asymptotics.
   } 
 \label{fig:reSigSsymTr}
\end{figure}

The real part of the frequency-dependent conductivity
$\re\sigma(\omega)$ is presented in figure~\ref{fig:reSigsymTr}, 
\ref{fig:reSigsymTrmuMq3}, \ref{fig:reSigSsymTr}, \ref{fig:reSigSdbiExp}. 
It shows the appearance
and growth of a gap as we increase the condensate $\dt^1_3$. 
The conductivity gap originates in a pseudo gap already present
right above $T_c$, as can be seen for example from the red curve 
in figure \ref{fig:reSigsymTr}. By pseudo gap we mean a well-defined gap in
the conductivity at low frequency in which the conductivity is not identically
zero \cite{Roberts:2008ns}. Both calculational prescriptions -- the adapted
symmetrized trace prescription and the expansion of the DBI action to fourth
order in the field strength -- yield qualitatively 
very similar results as can be seen by comparing figures \ref{fig:reSigSsymTr}
and \ref{fig:reSigSdbiExp}.
As a further distinct effect when using the adapted symmetrized
trace prescription, figure~\ref{fig:reSigsymTr}, 
\ref{fig:reSigsymTrmuMq3}, \ref{fig:reSigSsymTr} show 
prominent peaks which we interpret as mesonic excitations below. 
Increasing the quark mass $M_q$ from zero to a finite value, these
meson peaks become sharper, i.e. more quasiparticle-like.
This is reminiscent of results for condensed matter systems where prominent quasiparticle peaks appear (\eg \cite{RevModPhys.46.587}).
In figure \ref{fig:reSigsymTrmuMq3} at fixed $\mu/M_q=3$, the sharp resonances 
are also present inside the gap. 
In contrast to the adapted symmetrized trace prescription, from the 
expanded DBI action we obtain less prominent peaks as seen from 
figure \ref{fig:reSigSdbiExp}. In the conductivity obtained from the
DBI action expanded to fourth order, the peaks do not appear 
until we approach small temperatures. We expect that the terms
higher order in the field strength dominate the generation of the 
peaks and therefore the generation of the meson mass. If the higher
  order terms discussed here are included, we presume that the quasinormal
  modes which generate these peaks move closer to the real axis.

\FIGURE{
  \centering
  \psfrag{Resig}[c]{$\re \sigma$}
  \psfrag{w}{$\w$}
  \includegraphics[width=0.6\linewidth]{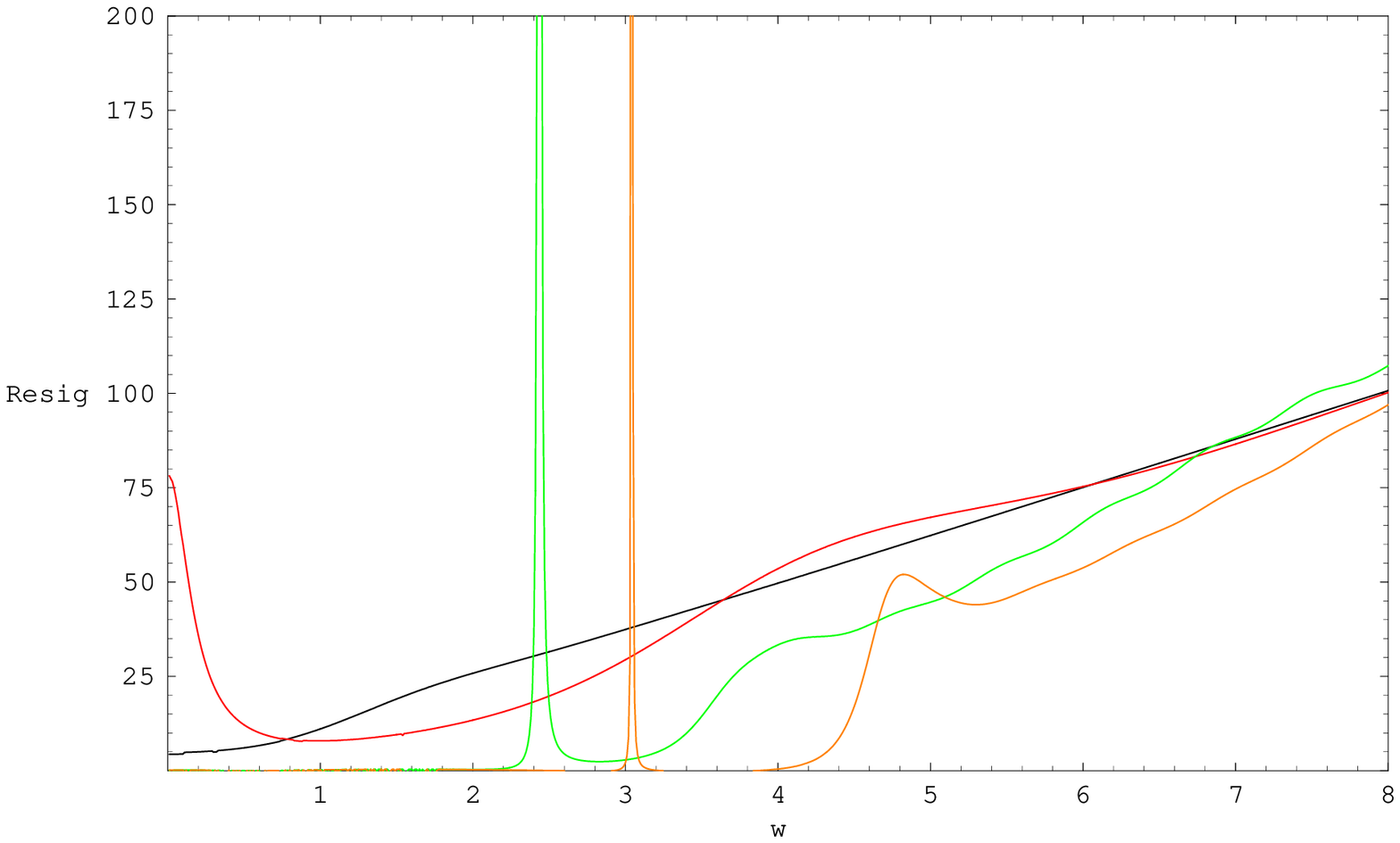}      
  \caption{Real part of conductivity, $\re \sigma,$ in units of
  $N_f N_c T/(16\pi)$ versus the dimensionless 
  frequency $\w=\omega/(2\pi T)$ for massive quarks at fixed $\mu/M_q=3$
  computed from the adapted symmetrized trace prescription.
  Distinct curves correspond to $T/T_c=10$ (black), $1$(red), 
  $0.6$ (green) and $0.5$ (orange). As in figure \ref{fig:reSigsymTr} a gap
  appears. Moreover, we observe a prominent peak inside the gap where the
    conductivity is approximately zero in the green and orange curves.
   } 
 \label{fig:reSigsymTrmuMq3}
}

\begin{figure}
  \centering
  \psfrag{ImsigS}[c]{$\w\im \sigma$}
  \psfrag{w}{$\w$}  
  \includegraphics[width=0.6\linewidth]{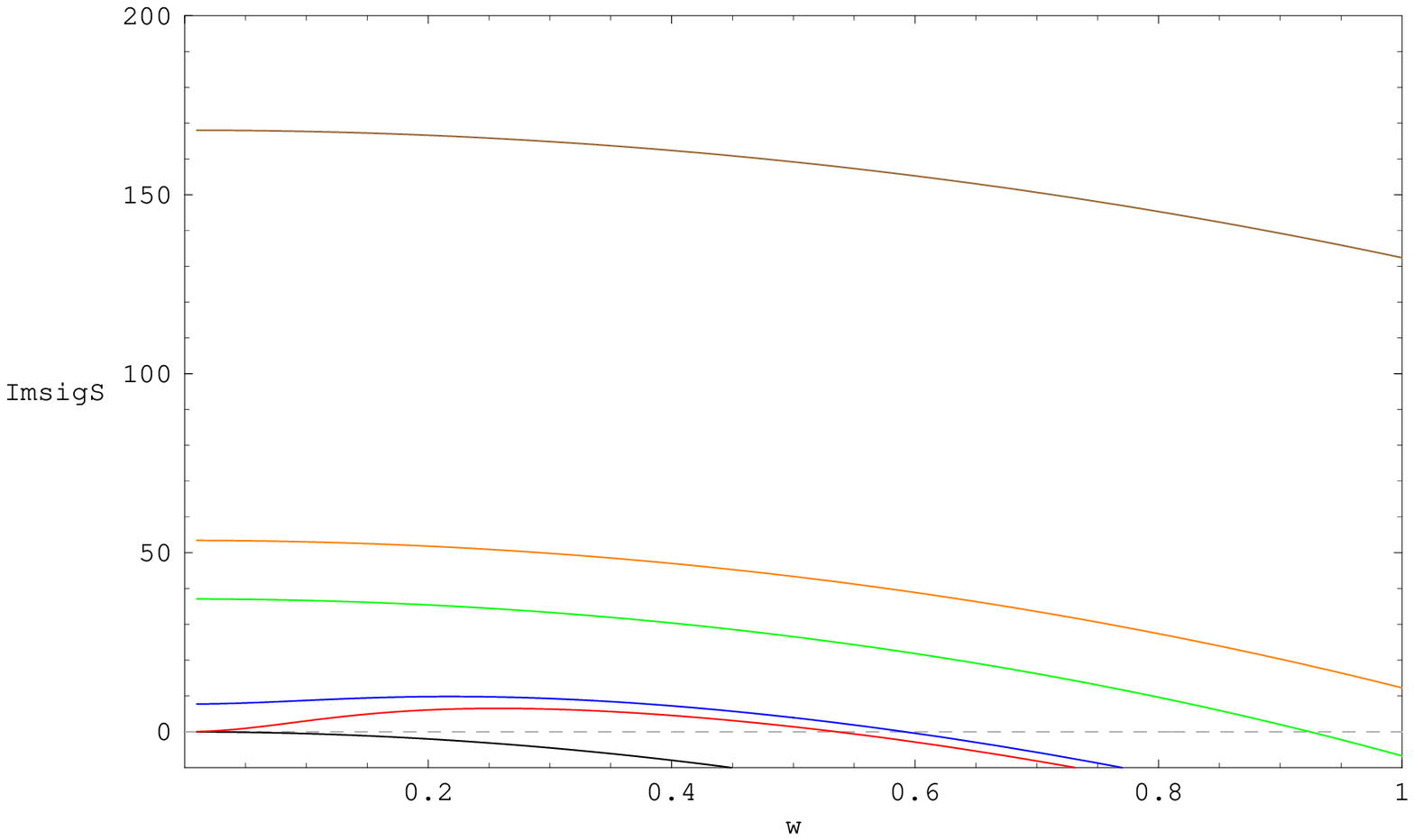}    
  \caption{Scaled imaginary part of conductivity
  $\w\im \sigma$ in units of
  $N_f N_c T/(16\pi)$ versus the dimensionless 
  frequency $\w=\omega/(2\pi T)$ for massless quarks
  computed from the adapted symmetrized trace prescription.
  Distinct curves correspond to $T/T_c=\infty$ (black), $1$ (red), $0.9$ (blue),
  $0.6$ (green), $0.5$ (orange) and $0.28$ (brown). This figure has been
  scaled to asymptote to a constant at $\w=0$. This constant
  determines the superconducting density $n_s$.
   } 
 \label{fig:imSigSsymTr}
\end{figure}

\begin{figure}
  \centering
  \psfrag{ResigS}{$\frac{\re \sigma}{4\pi\w}$}
  \psfrag{w}{$\w$}
  \includegraphics[width=0.6\linewidth]{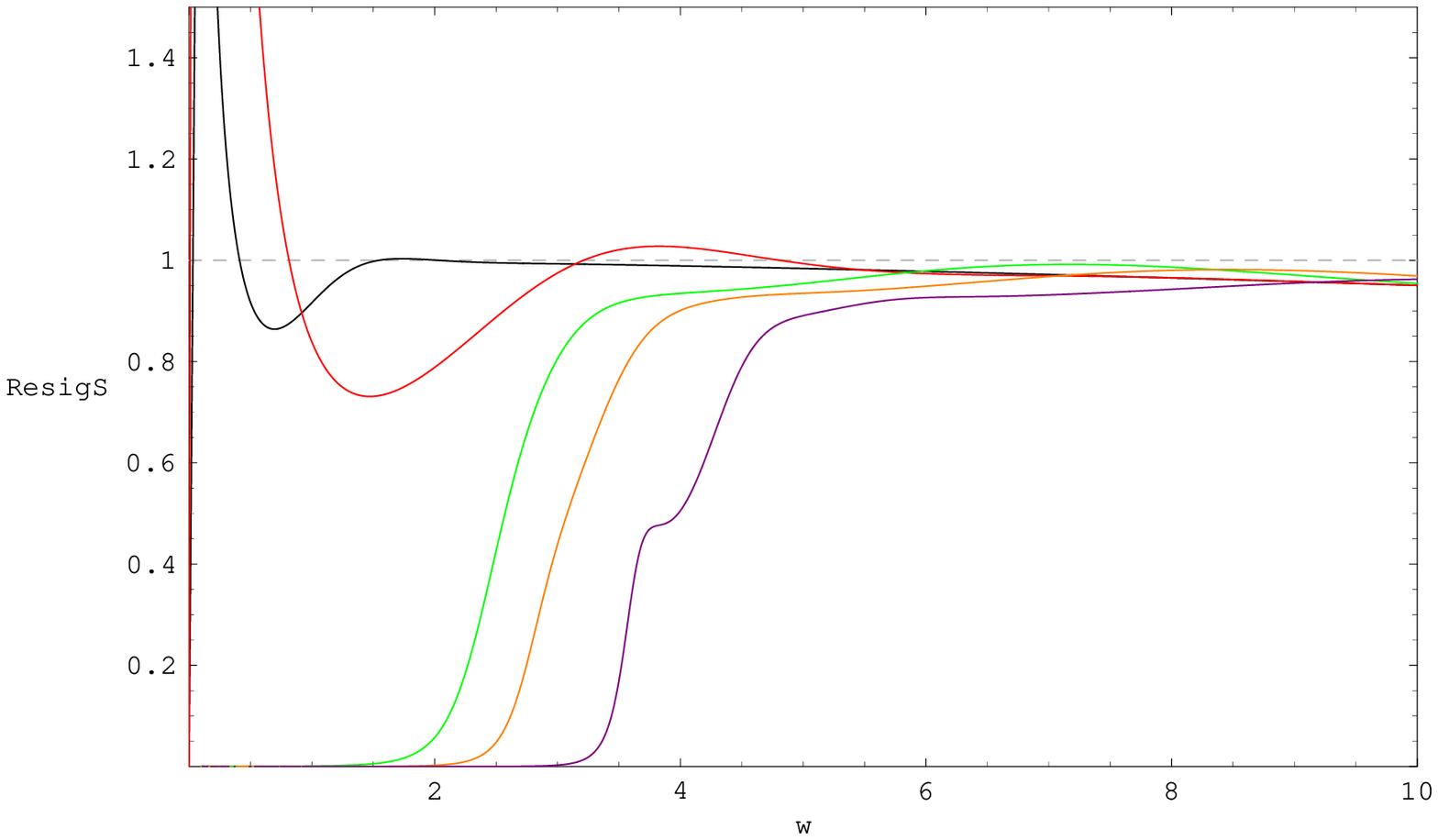}  
  \caption{Scaled real part of conductivity $\re \sigma/(4\pi \w)$ in units of
  $N_f N_c T/(16\pi)$ versus the dimensionless 
  frequency $\w=\omega/(2\pi T)$ for massless quarks
  computed from the expanded DBI action.
  Distinct curves correspond to $T/T_c=\infty$ (black), $1$(red), 
  $0.6$ (green), $0.5$ (orange) and $0.39$ (purple). This figure has been
  scaled to asymptote to a constant in order to show similarity
  to the lower dimensional cases computed from $AdS_4$ and 
  to show common asymptotics.
   } 
 \label{fig:reSigSdbiExp}
\end{figure}

Using the Kramers-Kronig relation, which connects the real and imaginary part of the
complex conductivity, we find a delta peak at $\omega=0$ in the real part of
the conductivity, $  \re\sigma(\omega)\sim\pi n_s\delta(\omega)$. The corresponding
$n_s/\omega$-behavior in the imaginary part is visualized in figure \ref{fig:imSigSsymTr}.
As expected from Ginzburg-Landau theory, our numerics show that the superconducting density
$n_s$ vanishes linearly at the critical temperature, $n_s\propto (1-T/T_c)$ for
$T\approx T_c$. This field theory definition of the superconducting density $n_s$ 
yields a quantity with the same linear scaling near $T_c$ as found in 
our bulk definition \eqref{eq:defds} of the superconducting density $\dt_s$. This confirms 
that these two quantities may be identified. Our numerics indeed indicate that $\dt_s$ and $n_s$ 
are proportional to each other. 

Note that for translation invariant systems at finite density,
there is a delta peak in the real part of the conductivity even in a normal
conducting phase since the charge carriers cannot lose their momentum. This peak
is called Drude peak. In our system, however, the charge carriers can dissipate
their momentum although our system is translation invariant
\cite{Karch:2007pd,Karch:2008uy}. The adjoint degrees of freedom can
transfer momentum at order $N_c^2$ while the fundamental degrees of freedom
only at order $N_c$. The adjoint degrees of freedom effectively act as a
heat sink into which the flavor fields can dissipate their momentum. Thus we
do not observe a Drude peak in our system.

\subsubsection{Spectral functions}
\label{sec:specFuncs}
\begin{figure}
  \centering
  \psfrag{ImG}[c]{$\mathcal{R}-\mathcal{R}_0$}
  \psfrag{w}{$\w$}  
  \includegraphics[width=0.6\linewidth]{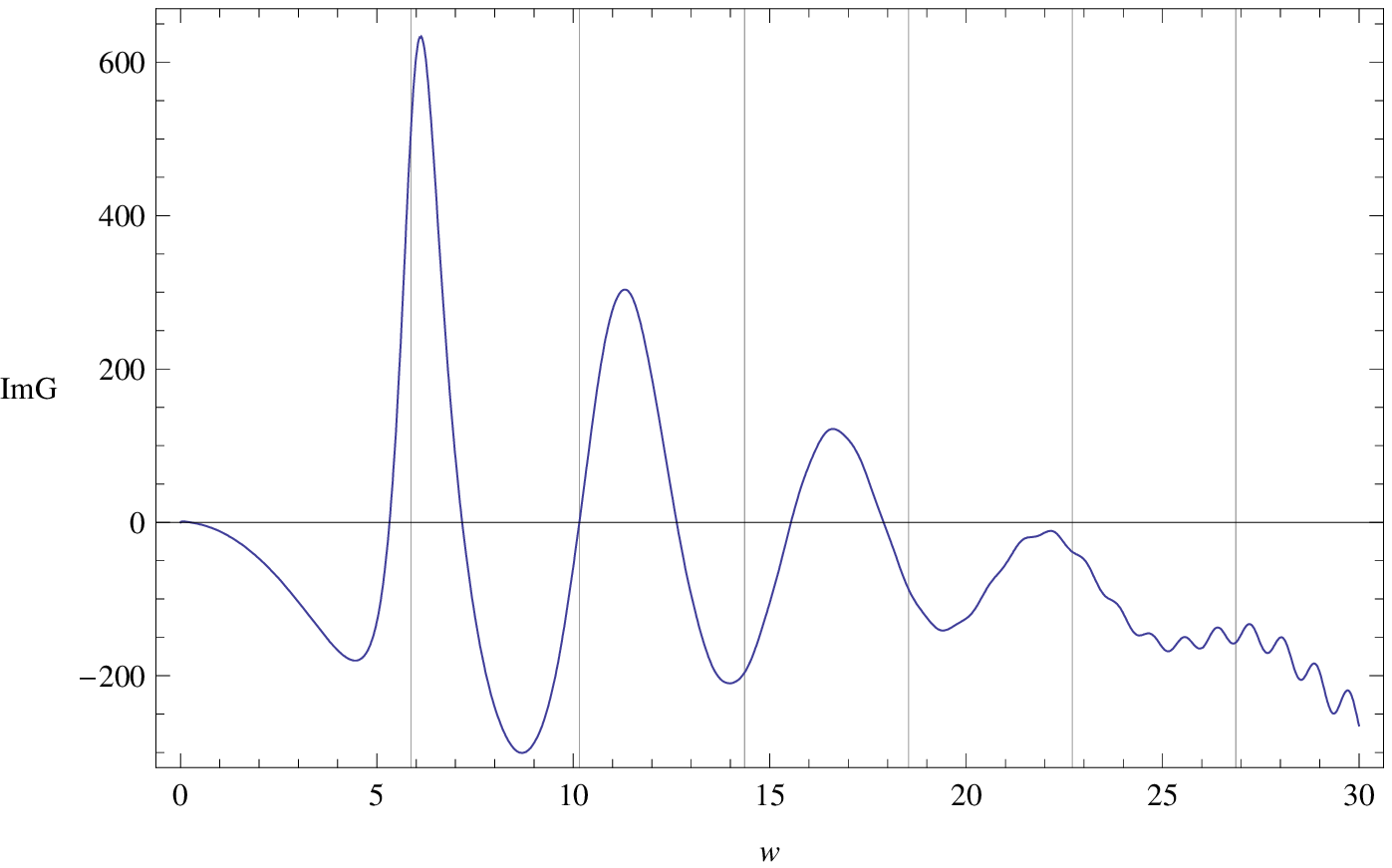}  
  \caption{Finite temperature part of the spectral function
    $\mathcal{R}-\mathcal{R}_0$ with $\mathcal{R}_0=4\pi\w^2$in units of 
  $N_f N_c T^2/8$ versus the dimensionless frequency $\w=\omega/(2\pi T)$ 
  at finite mass $m=2.842$ and chemical potential $\mut=3.483$. The grey lines
  correspond to the supersymmetric mass spectrum calculated in \cite{Kruczenski:2003be}.
   } 
 \label{fig:RhighmasssymTr}
\end{figure}

In order to identify the prominent peaks in the conductivity
in terms of field theory quantities as meson resonances, 
we examine the spectral functions  $\mathcal{R}=-2 \, \im G^R$
which are related to these conductivities by \eqref{eq:defsigma}.
Due to this relation the prominent peaks in the conductivity
correspond to the prominent peaks in the spectral function, see
figure \ref{fig:RsymTr}. As known from \cite{Myers:2007we,Erdmenger:2007ja}
the resonances appearing in the spectral functions of vector fields
in the bulk correspond to vector meson excitations in the 
dual field theory.
Moving to higher quark mass parameter $m=2.842$ and 
$\mut=3.483$ near the meson melting transition, we compare the 
resulting spectrum, see figure \ref{fig:RhighmasssymTr} to the 
supersymmetric mass formula obtained in \cite{Kruczenski:2003be}.
The prominent peaks clearly approach the supersymmetric line
spectrum from above. The same behavior was found in figure 7
of \cite{Erdmenger:2007ja}, where vector mesons
were considered also close to the meson melting transition. Note that the accuracy of
our numerics becomes insufficient at larger frequencies around $\w\approx 15$
\footnote{The frequency range of validity of our fluctuation numerics 
increases with quark mass. However, the numerics for the background
become worse at large quark masses.}.
\FIGURE{
  \centering
  \psfrag{R}{$\mathcal{R}$}
  \psfrag{w}{$\w$}
  \includegraphics[width=0.6\linewidth]{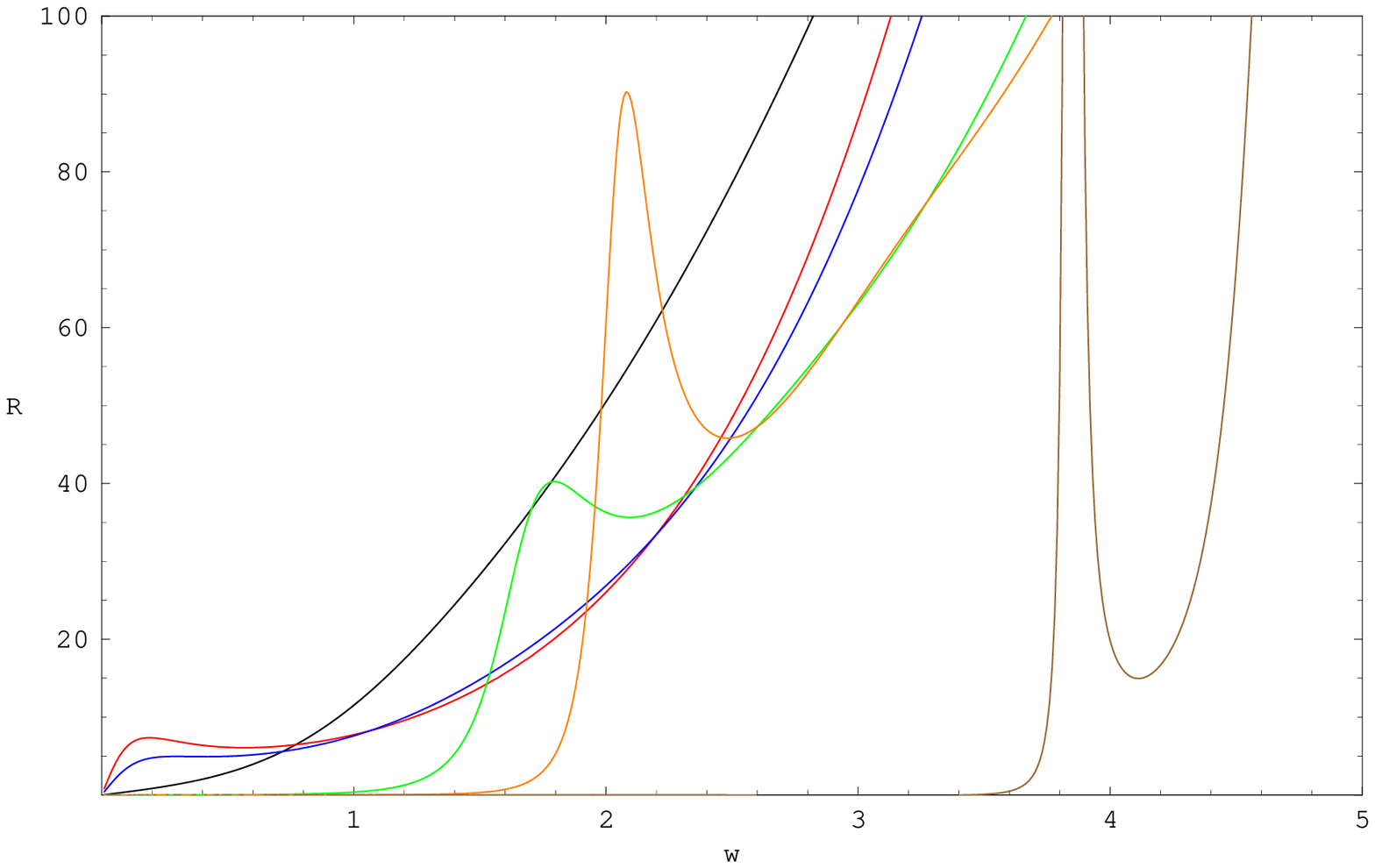}  
  \caption{The spectral function $\mathcal{R}$ in units of
  $N_f N_c T^2/8$ versus the dimensionless frequency $\w=\omega/(2\pi T)$ for
  massless quarks $M_q=0$. The distinct curves corresponds to different
  temperature (for color coding see figure~\ref{fig:imSigSsymTr}). When
  decreasing the temperature, peaks which correspond to mesonic quasiparticles
  emerge.} \label{fig:RsymTr}
}

\subsubsection{Comments on stability}
\label{sec:comments-stability}

\begin{figure}[h]
  \centering
  \psfrag{Xn0}[lc]{$n=0$}
  \psfrag{Xn1}[rc][rB]{$n=1$}
  \psfrag{Yn0}[rc][rB]{$n=0$}
  \psfrag{Yn1}[lc]{$n=1$}
  \psfrag{En0}[lb][rt]{$n=0$}
  \psfrag{Rew}{$\re\w$}
  \psfrag{Imw}{$\im\w$}
  \includegraphics[width=0.6\linewidth]{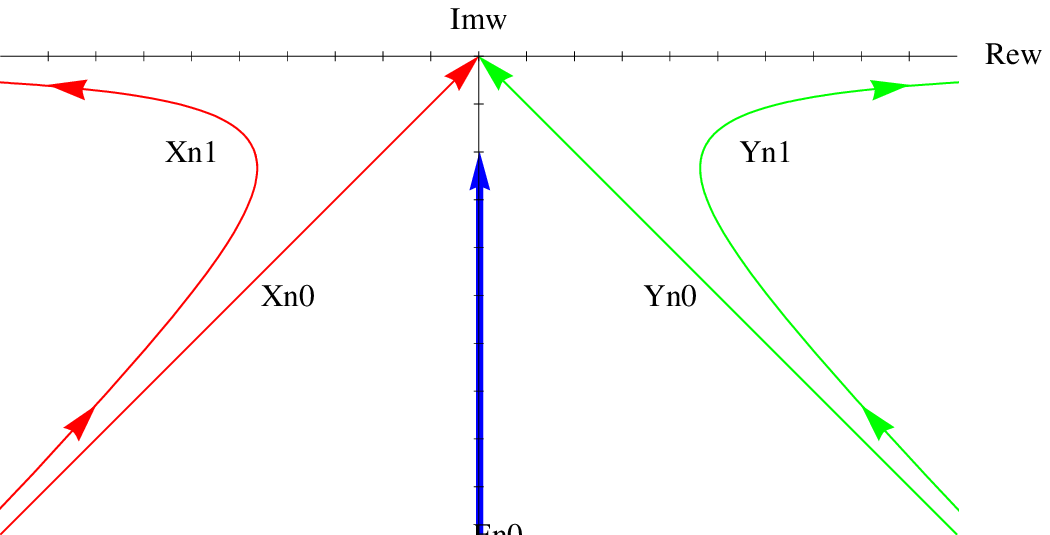}
  \caption{Movement of quasinormal modes under changes of the temperature $T$:
    The different colors indicate the different fluctuations $X$ (red),
    $Y$ (green) and $a^3_2$ (blue). The higher excitations of
    the fields $X$ and $Y$ behave as excitations with a non-zero quark mass. 
This indicates a dynamical generation of the meson mass.}
  \label{fig:qnm}
\end{figure}

As shown in figure~\ref{fig:qnm}, our setup is stable with respect to the
fluctuations $X$ and $Y$. Furthermore, figure~\ref{fig:qnm} shows that the
quasinormal modes of higher excitations $n>1$ move to larger frequencies and
closer to the real axis. This corresponds to the formation of 
stable massive mesons. Such a behavior is known in gauge/gravity duality for
mesons which are built from massive quarks (\eg\cite{Erdmenger:2007ja}).
Thus we observe a dynamical mass generation for the mesons as described in the following subsection. 

\subsubsection{Dynamical mass generation}
\label{sec:dynamicalMassGen}
In this section we discuss dynamical mas generation on the field theory
and on the gravity side. For this issue, it is important to distinguish
between the cases where the broken symmetry is global or local in the boundary
field theory.
\paragraph{Field theory observation}
As explained in section \ref{sec:dbi-action-equations}, the superconducting
condensate breaks the $U(1)_3$ symmetry spontaneously
in the field theory living on the AdS boundary. According
to the Goldstone theorem this generates one massless Nambu-Goldstone boson
in the boundary field theory. Let us now discuss the two different cases:
\begin{description}
\item[Local $U(1)_3$ symmetry:] If the broken symmetry is gauged, the
 Nambu-Goldstone boson is eaten by the gauge field $A^3$ charged under the
spontaneously broken $U(1)_3$ symmetry. In conventional superconductors this
mechanism gives mass to the photons which implies the Meissner-Ochsenfeld effect.
\item[Global $U(1)_3$ symmetry:] 
If the broken symmetry is global,  there is no
dynamical gauge field which eats the Nambu-Goldstone boson. Thus the
Nambu-Goldstone boson remains present in the spectrum.  Since this is the case
in our setup, we need to identify this Nambu-Goldstone boson in the spectrum.
In general, the Nambu-Goldstone boson corresponds to the phase of the condensate which parametrizes the the
coset space $U(1)_3/\Z_2$. In our setup, the fluctuations $X$ and $Y$ defined
above \eqref{eq:eomXY} are charged under the $U(1)_3$ symmetry such that the
Nambu-Goldstone boson can be found in these fluctuations (see
fig.~\ref{fig:qnm}). 
\end{description}

Although in our case the broken symmetry is global, we nevertheless observe
dynamical mass generation as see for instance in figures \ref{fig:RsymTr}
and~\ref{fig:qnm}. Therefore in our setup, a more
subtle mechanism then the ordinary Higgs mechanism generates the meson
masses dynamically. Let us explain this mechanism in the dual gravity setup.

\paragraph{Gravity explanation}
According to the AdS/CFT dictionary, a
global symmetry in the boundary field theory is mapped to a local symmetry in
the gravity theory. Thus the $SU(2)$ symmetry of the two D$7$-branes is local
in the gravity theory. In this paper we find new solutions to the equations of
motion which non-vanishing gauge field components $A_0^3$ and $A_3^1$. Since the
gravity action is unchanged and thus still $SU(2)$ invariant, this non-zero
components break the $SU(2)$ gauge symmetry spontaneously which generates in
total three Nambu-Goldstone bosons. They are eaten by the gauge fields which
then become massive. Let us now discuss the corresponding mass terms in more
detail.

First we consider a non-zero $SU(2)$ isospin chemical potential
$A^3_0$ which generates the mass term $(A^3_0)^2 (a^{1,2})^2$ for the
fluctuations $a^{1,2}$. This new term shifts the quadratic 
fluctuation term in the action, i.e. its energy by the value of the field $A^3_0$.
For the spectral function of the dual field theory this results in a 
shift of the meson resonance peaks, see \cite{Erdmenger:2007ja}.

Second we consider the non-zero gauge field component $A_3^1$ which
generates a mass term $(A_3^1)^2 (a^3)^2$ for the fluctuation $a^3$. We
expect that this mass term shifts the  quasinormal frequencies of this system in the complex
plane. That results in generation and shift of the resonance peaks in 
the spectral function as well as generation of the gap structure in the
conductivity. To make more precise statements, the spectrum of quasinormal
modes needs to be studied in presence of these fields.


\section{Meissner-Ochsenfeld effect}
\label{sec:meissner}
The Meissner effect is a distinct signature of conventional 
and unconventional superconductors. It is the phenomenon of 
expulsion of external magnetic fields. 
An induced current in the superconductor generates a magnetic
field counter-acting the external magnetic field~$H$. In AdS/CFT 
we are not able to observe the generation of counter-fields since
the symmetries on the boundary are always global. Nevertheless,
we can study their cause, i.e. the current induced in the superconductor.
As usual
\cite{Nakano:2008xc,Albash:2008eh,Maeda:2008ir,Hartnoll:2008kx,Wen:2008pb,Rebhan:2008ur}
the philosophy here is to weakly gauge the boundary theory afterwards.

In order to investigate how an external magnetic field influences our
p-wave superconductor, we have two choices. Either we introduce the
field along the spatial $z$-direction $H_3^3\tau^3$ or equivalently one along the $y$-direction
i.e. $H_2^3\tau^3$. Both are ``aligned''~with the spontaneously broken
$U(1)_3$-flavor direction.

As an example here we choose a non-vanishing $H_3^3\tau^3$. This requires
inclusion of some more non-vanishing field strength components in addition to those given
in equation~\eqref{eq:nonzeroF}. In particular we choose $A_0^3(\vrho,x),\,A_3^1(\vrho,x)$
and $A_2^3(x)=x^1\,H_3^3$ yielding the additional components\footnote{Close to
the phase transition, it is consistent to drop the dependence of the field
$H^3_3$ on $\vrho$. Away form the phase transition the $\vrho$ dependence must
be included. From the boundary asymptotics it will be possible to extract the
magnetic field and the magnetization of the superconductor.}
\begin{equation}
  \label{eq:additionalF}
  \begin{split}
    &H_3^3=F^3_{12}=-F^3_{21}=\del_1 A^3_2\, ,\\
    &F^1_{13}=-F^1_{31}=\del_1 A^1_3\,,\\
    &F^2_{23}=-F^2_{32}=\frac{\gamma}{\sqrt{\lambda}}A^3_2A^1_3\,,\\
    &F^3_{1 0}=-F^3_{01}=\del_1 A^3_0\,.
  \end{split}
\end{equation}
Recall that the radial AdS-direction is identified by
the indices either $\vrho$ or $4$.
Amending the DBI-action~\eqref{eq:DBI} with the additional
components~\eqref{eq:additionalF}, we compute the determinant in analogy to
equation~\eqref{eq:DBI}. We then choose to expand the new action to second
order in $F$, i.e. we only consider terms being at most quadratic in the
fields. This procedure gives the truncated DBI action 
\begin{equation}
   \label{eq:DBIwithB}
   \begin{split}
   S_{\text{DBI}}=&-T_{D7}N_f\int\!\dd^8\xi\:\sqrt{-G}\Bigg[1+\frac{(2\pi\alpha')^2}{2} \Big(G^{00}G^{33} (F^2_{03})^2
    +G^{33}G^{44} (F^1_{\vrho 3})^2+G^{00}G^{44}(F^3_{\vrho 0})^2\\ 
   &+(G^{33})^2(F_{13}^1)^2+(G^{33})^2(F_{23}^2)^2+G^{00}G^{33}(F_{10}^3)^2 
    +(G^{33})^2(F_{12}^3)^2\Big)+\cdots\Bigg]\,.
  \end{split}
 \end{equation}
Respecting the symmetries and 
variable dependencies in our specific system, this can be written as
\begin{equation}
   \label{eq:DBIAwithB}
   \begin{split}
   S_{\text{DBI}}=&-T_{D7}N_f\int\!\dd^8\xi\:\sqrt{-G}
   \Bigg[
   1+\frac{1}{2} \Big(G^{00}G^{33} (\del_{\bar 1}\tilde A_0^3)^2+G^{33}G^{44} (\del_\vrho \tilde A_3^1)^2+ 
   G^{00}G^{44}(\del_{\vrho}\tilde A_0^3)^2\\ 
   &+(G^{33})^2(\del_{\bar 1} \tilde A_3^1)^2
   +(G^{33})^2 (\bar H_3^3)^2+G^{00}G^{33}
   \frac{{\vrho_H}^4\gamma^2}{(2\pi\alpha')^2\lambda}
   (\tilde A_0^3\tilde A_3^1)^2\\  
   &+(G^{33})^2\frac{{\vrho_H}^4\gamma^2}{(2\pi\alpha')^2\lambda}
    (\tilde A_3^1 \bar H_3^3)^2{\bar x}^2\Big)+\cdots\Bigg]\,,
  \end{split}
 \end{equation}
with the convenient redefinitions
\begin{equation}
 \label{eq:redefinitions}
 \tilde A=\frac{2\pi\alpha'}{\vrho_H} A\, ,\quad x=\vrho_H \bar x\, ,\quad
 \bar H_3^3=2\pi\alpha' H_3^3\,, \quad \vrho=\vrho_H\, \rho\,.
\end{equation}
Rescaling the $\bar x$-coordinate once more
\begin{equation}
\tilde x=\sqrt{\frac{\bar H_3^3 {\vrho_H}^2 \gamma}{2\pi\alpha'\sqrt{\lambda}}}\,\bar x\,,
\end{equation}
the equations of motion derived from the action \eqref{eq:DBIAwithB} take a simple form
\begin{equation}
 \label{eq:eomAB}
 \begin{split}
  0=&\del_\rho^2\tilde A_0^3+\frac{\del_\rho\left[\sqrt{-G} G^{00} G^{44}\right]}{\sqrt{-G} G^{00} G^{44}}
  \del_\rho \tilde A_0^3+ \frac{\gamma \tilde H_3^3}{2\sqrt{2}\pi}\frac{G^{33}}{G^{44}}
  \del_{\tilde x}\tilde A_0^3-\frac{\gamma^2}{2\pi^2}\frac{G^{33}}{G^{44}}
  \tilde A_0^3(\tilde A_3^1)^2\, ,\\
  0=&\del_\rho^2\tilde A_3^1+\frac{\del_\rho\left[\sqrt{-G} G^{33} G^{44}\right]}{\sqrt{-G} G^{33} G^{44}}
  \del_\rho \tilde A_3^1 + \frac{\gamma \tilde H_3^3}{2\sqrt{2}\pi}\frac{G^{33}}{G^{44}}
  \left[\del^2_{\tilde x}\tilde A_3^1-{\tilde x}^2\tilde A_3^1\right]-\frac{\gamma^2}{2\pi^2}\frac{G^{00}}{G^{44}}(\tilde A_0^3)^2\tilde A_3^1 \, .
 \end{split}
\end{equation}
Here all metric components are understood to be evaluated at $R=1$ and $\vrho\to\rho$.

\begin{figure}
  \centering
  \psfrag{B}[r]{$\frac{H_3^3}{d_0^3}$}
  \psfrag{T}[tr]{$\frac{T}{{d_0^3}^{1/3}}$}
  \includegraphics[width=0.6\linewidth]{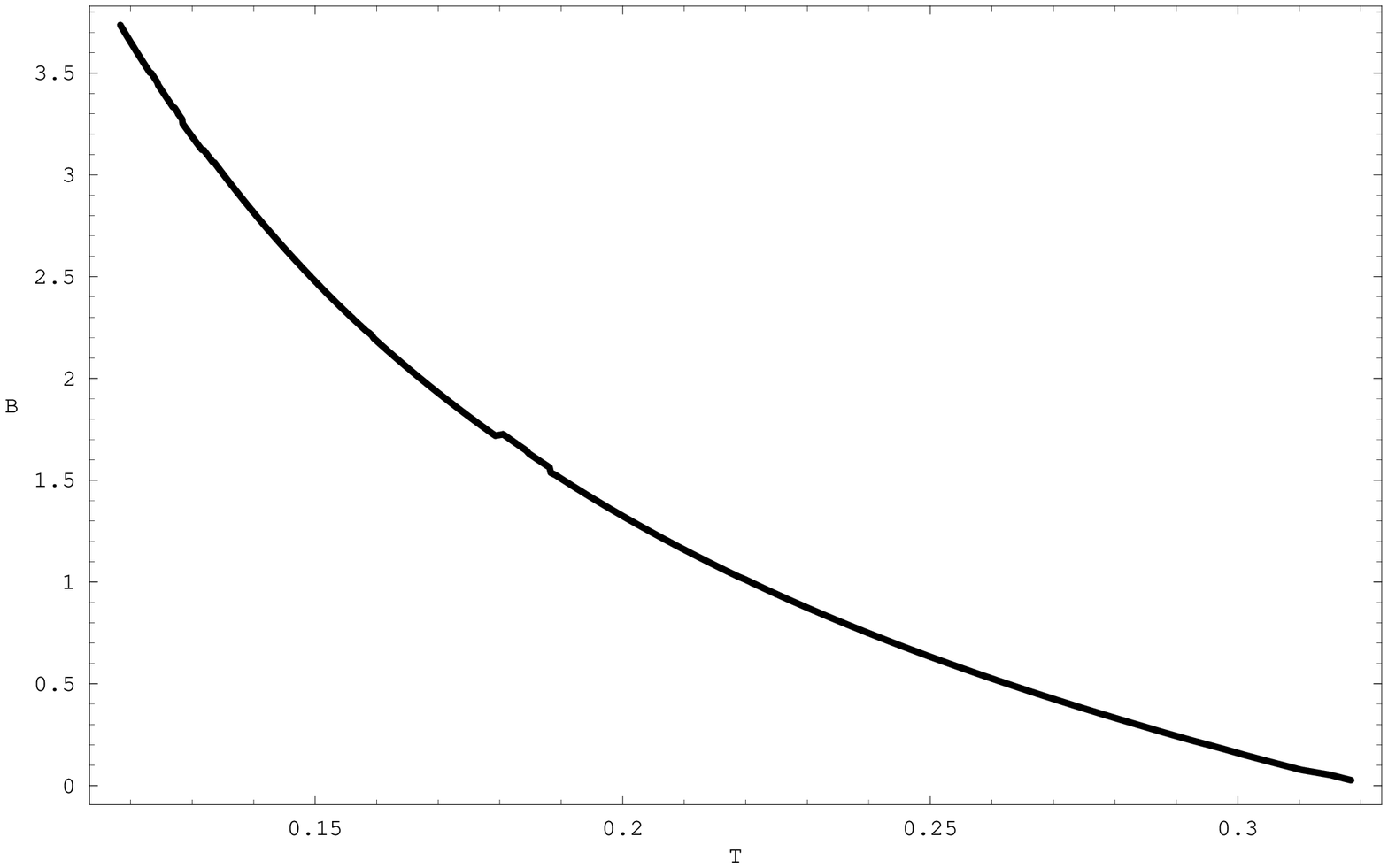}
  \caption{\label{fig:criticalBT}
    The line of critical magnetic field versus critical temperature. Below
    this line the external magnetic field coexists with the superconducting
    condensate. Above the line the superconducting condensate vanishes.
    We set the spatial position arbitrarily to $\tilde x=0.1$ since the critical
    line does not depend on $\tilde x$.
  }
\end{figure}

We aim at decoupling and solving the system of partial differential
equations~\eqref{eq:eomAB} by the product ansatz
\begin{equation}\label{eq:prodAnsatz}
\tilde A_3^1(\rho,\tilde x)= \mathfrak{g}(\rho)\, \mathfrak{f}(\tilde x)\,.
\end{equation}
For this ansatz to work, we need to make two assumptions: First we assume 
that $\tilde A_0^3$ is constant in $\tilde x$. Second we assume that $A_3^1$
is small, which clearly is the case near the transition $T\to T_c$. Our second
assumption prevents $A_0^3$ from receiving a dependence on $\tilde x$ 
through its coupling to $A_3^1(\rho,\tilde x)$. These assumptions allow to
write the second equation in \eqref{eq:eomAB} as
\begin{equation}
 \label{eq:eomg}
 \begin{split}
  0=&\del_\rho^2 \mathfrak{g}(\rho)+\frac{\del_\rho\left[\sqrt{-G} G^{33} G^{44}\right]}{\sqrt{-G} G^{33} G^{44}}
  \del_\rho \mathfrak{g}(\rho)-\frac{\gamma^2}{2\pi^2}\frac{G^{00}}{G^{44}}(\tilde A_0^3)^2 \mathfrak{g}(\rho)\\   
  &+ \frac{\gamma \tilde H_3^3}{2\sqrt{2}\pi}\frac{G^{33}}{G^{44}}
  \,\mathfrak{g}(\rho)\,\frac{\del^2_{\tilde x} \mathfrak{f}
  (\tilde x)-{\tilde x}^2 \mathfrak{f}(\tilde x)}{\mathfrak{f}
  (\tilde x)}\, .
 \end{split}
\end{equation}
All terms but the last one are independent of $\tilde x$, so the product 
ansatz \eqref{eq:prodAnsatz} is consistent only if 
\begin{equation}\label{eq:eomf}
\frac{\del^2_{\tilde x} \mathfrak{f}(\tilde x)-{\tilde x}^2 \mathfrak{f}(\tilde x)}{\mathfrak{f}(\tilde x)}=C\,,
\end{equation}
where $C$ is a constant. The differential equation \eqref{eq:eomf} has a
particular solution if $C=-(2n+1)\,, n\in \mathbb{N}$. The solutions for $\mathfrak{f}(\tilde x)$
are Hermite functions 
\begin{equation}
\mathfrak{f}_n(\tilde x)=\frac{e^{-\frac{|\tilde x|^2}{2}}}{\sqrt{n!2^n\sqrt{\pi}}}H_n(\tilde x)\,,\,\, 
 H_n(\tilde x)=(-1)^n e^{\frac{|\tilde x|^2}{2}}\frac{d^n}{dx^n} e^{-\frac{|\tilde x|^2}{2}},
\end{equation}
which have Gaussian decay at large $|\tilde x|\gg1$. 
Choosing the lowest solution with $n=0$ and $H_0=1$, which has no nodes, is most likely to give
the configuration with lowest energy content. 
So the system we need to solve is finally given by
\begin{equation}
 \label{eq:eomgA}
 \begin{split}
  0=\del_\rho^2\tilde A_0^3&+\frac{\del_\rho\left[\sqrt{-G} G^{00} G^{44}\right]}{\sqrt{-G} G^{00} G^{44}}
  \del_\rho \tilde A_0^3\frac{\gamma^2}{2\pi^2}\frac{G^{33}}{G^{44}}
 \tilde A_0^3(\mathfrak{f}_0(\tilde x)\mathfrak{g}(\rho))^2\, , \\ 
   0=\del_\rho^2 \mathfrak{g}&+\frac{\del_\rho\left[\sqrt{-G} G^{33} G^{44}\right]}{\sqrt{-G} G^{33} G^{44}}
  \del_\rho \mathfrak{g}-\frac{\gamma^2}{2\pi^2}\frac{G^{00}}{G^{44}}(\tilde A_0^3)^2 \mathfrak{g}
  -\frac{\gamma \tilde H_3^3}{2\sqrt{2}\pi}\frac{G^{33}}{G^{44}}\,\mathfrak{g}\, . 
 \end{split}
\end{equation}
Asymptotically near the horizon the fields take the form
\begin{equation}\label{eq:horAs}
 \begin{aligned}
  \,&\At_0^3 &&=          &&+a_2 (\rho-1)^2                        &&+\mathcal{O}((\rho-1)^{3})\, ,\\
  \,&\mathfrak{g}&&=b_0    &&+\frac{b_0 H_3^3}{4} (\rho-1)^2&&+\mathcal{O}((\rho-1)^{-3})\,,
 \end{aligned}
\end{equation}
while at the boundary we obtain
\begin{equation}\label{eq:bdyAs}
 \begin{aligned}
  \,&\tilde A_0^3 &&=\tilde \mu &&+\frac{\tilde d_0^3}{\rho^{2}}       &&+\mathcal{O}(\rho^{-4})\, ,\\
  \,&\mathfrak{g}&&=               &&+\frac{\tilde d_3^1}{\rho^{2}}&&+\mathcal{O}(\rho^{-4})\, .
 \end{aligned}
\end{equation}
We succeed in finding numerical solutions $\mathfrak{g}(\rho)$ and
$A_0^3(\rho)$ to the set of equations~\eqref{eq:eomgA} obeying the asymptotics
given by equations \eqref{eq:horAs} and \eqref{eq:bdyAs}. These numerical
solutions are used to approach the phase transition from the
superconducting phase by increasing the magnetic field. We map out the line of critical
temperature-magnetic field pairs in figure \ref{fig:criticalBT}. In this way
we obtain a phase diagram displaying the Meissner effect. The critical line in
figure \ref{fig:criticalBT} separates the phase with and without
superconducting condensate $\dt_3^1$. 

We emphasize that this is a background  calculation involving no fluctuations.
Complementary to the procedure described above we also confirmed the phase diagram using  
the instability of the normal phase against fluctuations. Starting at large
magnetic field and vanishing condensate $\tilde d_3^1$, we determine for a
given magnetic field $H_3^3$ the temperature $T_c(H_3^3)$ at which the
fluctuation $a_1^3$ becomes unstable. That instability signals the
condensation process into the superconducting phase.

The presence of the coexistence phase below the critical line, where the system
is still superconducting despite the presence of an external magnetic field,
is the signal of the Meissner effect in the case of a global symmetry
considered here. If we now weakly gauged the flavor symmetry at the boundary,
the superconducting current $J_3^1$ would generate a magnetic field opposite
to the external field. Thus the phase observed is a necessary condition in the
case of a global symmetry for finding the standard Meissner effect when
gauging the symmetry.


\section{Discussion and Outlook}
\label{sec:discussion}
We found a holographic realization of superconductivity in the context
of gauge/gravity duality with flavor at finite isospin chemical
potential, for which the field theory action is known explicitly. The
condensation process corresponds to a recombination of strings which
leads to a thermodynamically favored configuration. On the field theory
side, this new state may be interpreted as a $\rho$ meson superfluid. We
show that the superconducting phase persists at finite quark mass and
find a holographic Meissner--Ochsenfeld effect. From a technical point
of view, we compare two different approaches for evaluating the
non-Abelian DBI action involved (expansion to fourth order and a new
modified trace evaluation procedure which we propose) and find the same
physical results in both approaches, at least qualitatively.

For the future, we expect that a holographic s-wave superconductor may
be constructed by applying the methods presented to the transverse
scalars in similar holographic setups. It will also be interesting to
apply our methods to the configuration discussed in \cite{Myers:2008me} in
order to find a purely fermionic superconductor.


\section*{Acknowledgements}
We are grateful to T. Dahm, S.~Gubser, S.~Hartnoll, C.~Herzog, K.~Landsteiner,
R.~Meyer, A. O'Bannon, S.~Pufu, F.~Rust and E. Tsatis for discussions. This work was supported in
part by  {\it The Cluster of Excellence for Fundamental Physics - Origin and
  Structure of the Universe}.

\begin{appendix}
\section{Expansion of the DBI action to fourth order}
\label{sec:expans-fourth-order}
In this section we give some lengthy equations needed to calculate the
expanded DBI action and the resulting equations of motion (see section \ref{sec:expansion-dbi-action}).

The $\calt_i$ which determine the terms in expanded DBI action at $i-th$ order
in the field strength are given by
\begin{equation}
  \label{eq:calt}
  \begin{split}
    \calt_2=&(2\pi\alpha')^2\left[G^{00}G^{44}\left(F_{\vrho 0}^3\right)^2+G^{33}G^{44}\left(F_{\vrho
        3}^1\right)^2+G^{00}G^{33}\left(F_{03}^2\right)^2\right]\\
  =&-\frac{2\ft
    H}{f^2}(\del_\rho\At_0^3)^2+\frac{2H}{\ft}(\del_\rho\At_3^1)^2-\frac{2\gamma^2}{\rho^4f^2\pi^2}(\At_0^3\At_3^1)^2\,,\\
    \calt_4=&(2\pi\alpha')^4\Big[\left(G^{00}G^{44}\right)^2\left(F_{\vrho
        0}^3\right)^4+\left(G^{33}G^{44}\right)^2\left(F_{\vrho
        3}^1\right)^4+\left(G^{00}G^{33}\right)^2\left(F_{03}^2\right)^4\\
    &+\frac{2}{3}G^{00}G^{33}\left(G^{44}\right)^2\left(F_{\vrho
        0}^3\right)^2\left(F_{\vrho 3}^1\right)^2+\frac{2}{3}\left(G^{00}\right)^2G^{33}G^{44}\left(F_{\vrho
        0}^3\right)^2\left(F_{0 3}^2\right)^2\\
    &+\frac{2}{3}G^{00}\left(G^{33}\right)^2G^{44}\left(F_{\vrho
        3}^1\right)^2\left(F_{0 3}^2\right)^2\Big]\\
    =&\frac{4\ft^2H^2}{f^4}(\del_\rho\At_0^3)^4+\frac{4H^2}{\ft^2}(\del_\rho\At_3^1)^4+\frac{4\gamma^4}{\rho^8f^4\pi^4}(\At_0^3\At_3^1)^4-\frac{8H^2}{3f^2}(\del_\rho\At_0^3)^2(\del_\rho\At_3^1)^2\\
    &+\frac{8\ft H\gamma^2}{3\rho^4f^4\pi^2}(\del_\rho\At_0^3)^2(\At_0^3\At_3^1)^2-\frac{8H\gamma^2}{3\rho^4\ft
    f^2\pi^2}(\del_\rho\At_3^1)^2(\At_0^3\At_3^1)^2\,,\\
  \end{split}
\end{equation}
with
\begin{equation}
  \label{eq:2}
  H=\frac{1-\chi^2}{1-\chi^2+\rho^2(\del_\rho\chi)^2}\,.
\end{equation}
Using the Euler-Lagrange equation the resulting equations of motion are\small
\begin{equation}
    \label{eq:eomexpand}
    \begin{split}
      &\del_\rho\Bigg[\sqrt{-G}\Bigg(\frac{2\ft H}{f^2}(\del_\rho
      \At_0^3)+\frac{2\ft^2H^2}{f^4}(\del_\rho\At_0^3)^3-\frac{2H^2}{3f^2}(\del_\rho\At_3^1)^2(\del_\rho\At_0^3)+\frac{2\ft\gamma^2H}{3f^4\rho^4\pi^2}(\del_\rho\At_0^3)(\At_0^3\At_3^1)^2\Bigg)\Bigg]\\
      &=\frac{2\gamma^2\sqrt{-G}}{\pi^2\rho^4f^2}\Bigg[\At_0^3(\At_3^1)^2+\frac{\gamma^2}{\pi^2\rho^4f^2}(\At_0^3)^3(\At_3^1)^4+\frac{\ft 
        H}{3f^2}(\del_\rho\At_0^3)^2\At_0^3(\At_3^1)^2-\frac{H}{3\ft}(\del_\rho\At_3^1)^2\At_0^3(\At_3^1)^2\Bigg]\,,\\
      &\del_\rho\Bigg[\sqrt{-G}\Bigg(\frac{2H}{\ft}(\del_\rho\At_3^1)-\frac{2H^2}{\ft^2}(\del_\rho\At_3^1)^3+\frac{2H^2}{3f^2}(\del_\rho\At_0^3)^2(\del_\rho\At_3^1)+\frac{2\gamma^2H}{3\pi^2\rho^4f^2\ft}(\del_\rho\At_3^1)(\At_0^3\At_3^1)^2\Bigg)\Bigg]\\
      &=-\frac{2\gamma^2\sqrt{-G}}{\pi^2\rho^4f^2}\Bigg[(\At_0^3)^2\At_3^1+\frac{\gamma^2}{\pi^2\rho^4f^2}(\At_0^3)^4(\At_3^1)^3+\frac{\ft
        H}{3f^2}(\del_\rho\At_0^3)^2(\At_0^3)^2\At_3^1-\frac{H}{3\ft}(\del_\rho\At_3^1)^2(\At_0^3)^2\At_3^1\Bigg]\,,\\
      &\del_\rho\Bigg[\frac{\rho^5f\ft (1-\chi^2)(\del_\rho\chi)}{\sqrt{1-\chi^2+\rho^2(\del_\rho\chi)^2}}\Bigg(1+\frac{\calt_2}{2}-\frac{\calt_4}{8}-\frac{\vrho_H^2}{R^2}\frac{2\rho^2(1-\chi^2)}{1-\chi^2+\rho^2(\del_\rho\chi)^2}
      \left[\frac{1}{2}\frac{\del\calt_2}{\del
          G^{44}}-\frac{1}{8}\frac{\del\calt_4}{\del
          G^{44}}\right]\Bigg)\Bigg]\\
      &=-\frac{\rho^3f\ft\chi}{\sqrt{1-\chi^2+\rho^2(\del_\rho\chi)^2}}\Bigg[
      \left(3(1-\chi^2)+2\rho^2(\del_\rho\chi)^2\right)
      \left(1+\frac{\calt_2}{2}-\frac{\calt_4}{8}\right)\\
      &\quad+\frac{\vrho_H^2}{R^2}\frac{2\rho^4(1-\chi^2)(\del_\rho\chi)^2}{1-\chi^2+\rho^2(\del_\rho\chi)^2}\left(\frac{1}{2}\frac{\del\calt_2}{\del G^{44}}-\frac{1}{8}\frac{\del\calt_4}{\del G^{44}}\right)\Bigg]\,,
    \end{split}
  \end{equation}\normalsize
with
\begin{equation}
  \label{eq:3}
  \begin{split}
    \frac{\del\calt_2}{\del
      G^{44}}=&\frac{2R^2}{\vrho_H^2\rho^2}\left[-\frac{\ft}{f^2}(\del_\rho\At_0^3)^2+\frac{1}{\ft}(\del_\rho\At_3^1)^2\right]\,,\\
    \frac{\del\calt_4}{\del G^{44}}=&\frac{2R^2}{\vrho_H^2\rho^2}
    \Bigg[\frac{4\ft^2H}{f^4}(\del_\rho\At_0^3)^4+\frac{4H}{\ft^2}(\del_\rho\At_3^1)^4-\frac{8H}{3f^2}(\del_\rho\At_0^3)^2(\del_\rho
      \At_3^1)^2\\
      &+\frac{4\gamma^2\ft}{3\rho^4f^4\pi^2}(\del_\rho\At_0^3)^2(\At_0^3\At_3^1)^2-\frac{4\gamma^2}{3\rho^4\ft  f^2\pi^2}(\del_\rho\At_3^1)^2(\At_0^3\At_3^1)^2\Bigg]\,.
  \end{split}
\end{equation}

\section{Fluctuations}
\label{sec:fluctuationsapp}

\subsection{Adapted symmetrized trace prescription}
\label{sec:adapt-symm-trace}

In this section we present the elements of the inverse background tensor
$\G^{\mu\nu}$ defined in section \ref{sec:adopt-symm-trace}. The diagonal
elements are given by
\begin{equation}
  \label{eq:inverseMetric}
  \begin{split}
    &\G^{00}=G^{00}\frac{1+G^{33}G^{44}(\del_\rho \At^1_3)^2}{N}\,,\quad \G^{33}=G^{33}\frac{1+G^{00}G^{44}(\del_\rho \At^3_0)^2}{N}\,,\\
    &\G^{44}=G^{44}\frac{1-\frac{2\gamma^2}{\pi^2\rho^4f^2}(\At^1_3\At^3_0)^2}{N}\,,
  \end{split}
\end{equation}
where the other diagonal elements are just $\G^{ii}=G^{ii}$. The
off-diagonal elements read
\begin{equation}
\begin{aligned}
    \G^{34}=&G^{44}\frac{G^{33}(\del_\rho
      \At^1_3)\tau^1-\frac{2\sqrt{2}\gamma}{\pi\rho^4f^2}\frac{R^2}{\vrho_H^2}\At^1_3\At^3_0(\del_\rho
      \At^3_0)\tau^2\tau^3}{N}\,,\\
    \G^{43}=&G^{44}\frac{-G^{33}(\del_\rho
      \At^1_3)\tau^1-\frac{2\sqrt{2}\gamma}{\pi\rho^4f^2}\frac{R^2}{\vrho_H^2}\At^1_3\At^3_0(\del_\rho
      \At^3_0)\tau^2\tau^3}{N}\\
    \G^{03}=&\frac{4R^2}{\vrho_H^2\rho^4f^2}\frac{\frac{\sqrt{2}\gamma}{2\pi}\At^1_3\At^3_0\tau^2+\frac{R^2}{\vrho_H^2}G^{44}(\del_\rho
      \At^1_3)(\del_\rho \At^3_0)\tau^1\tau^3}{N}\,,\\
    \G^{30}=&\frac{4R^2}{\vrho_H^2\rho^4f^2}\frac{-\frac{\sqrt{2}\gamma}{2\pi}\At^1_3\At^3_0\tau^2+\frac{R^2}{\vrho_H^2}G^{44}(\del_\rho \At^1_3)(\del_\rho \At^3_0)\tau^1\tau^3}{N}\\
    \G^{04}=&G^{00}G^{44}\frac{(\del_\rho\At^3_0)\tau^3-\frac{\sqrt{2}\gamma}{2\pi}\frac{\vrho_H^2}{R^2}G^{33}\At^1_3\At^3_0(\del_\rho
      \At^1_3)\tau^1\tau^2}{N}\,,\\
    \G^{40}=&G^{00}G^{44}\frac{-(\del_\rho\At^3_0)\tau^3-\frac{\sqrt{2}\gamma}{2\pi}\frac{\vrho_H^2}{R^2}G^{33}\At^1_3\At^3_0(\del_\rho
      \At^1_3)\tau^1\tau^2}{N}\,.
  \end{aligned}
\end{equation}
The denominator $N$ is given by
\begin{equation}
  \label{eq:inverseMatricdenom}
    N=1+G^{33}G^{44}(\del_\rho
    \At^1_3)^2+G^{00}G^{44}(\del_\rho
    \At^3_0)^2-\frac{2\gamma^2}{\pi^2\rho^4f^2}(\At^1_3\At^3_0)^2\,.
\end{equation}

\subsection{Expansion of the DBI action to fourth order}
\label{sec:expas-fourth-order}

In this section we calculate the quadratic action for the fluctuations using the
expanded DBI action as explained in section \ref{sec:expansion-dbi-action-1}.
Since the fluctuation $a_2^3$ decouples from every other fluctuation we can
write down an effective Lagrangian for this fluctuation. This Lagrangian is
given by 
\small
\begin{equation}
  \label{eq:effLa32}
  \begin{split}
  \call_{\text{eff}}^{a_2^3}=&\sqrt{-G}\:\str\Bigg[\frac{1}{2}G^{22}G^{44}(\check
  F_{24}^3)^2(\sigma^3)^2+\frac{1}{2}G^{00}G^{22}(\check
  F_{02}^3)^2(\sigma^3)^2+\frac{1}{2}G^{22}G^{33}(\check
  F_{23}^2)^2(\sigma^2)^2\\
  &+\frac{1}{4}G^{33}G^{44}(F_{34}^1)^2
  \left[G^{00}G^{22}(\check
    F_{02}^3)^2(\sigma^3\sigma^1)^2-G^{22}G^{44}(\check
    F_{24}^3)^2(\sigma^3\sigma^1)^2-G^{22}G^{33}(\check F_{23}^2)^2(\sigma^2\sigma^1)^2\right]\\
  &+\frac{1}{4}G^{00}G^{33}(F_{03}^2)^2
  \left[G^{22}G^{44}(\check
    F_{24}^3)^2(\sigma^3\sigma^2)^2-G^{00}G^{22}(\check
    F_{02}^3)^2(\sigma^3\sigma^2)^2-G^{22}G^{33}(\check
    F_{23}^2)^2(\sigma^2)^4\right]\\
  &+\frac{1}{4}G^{00}G^{44}(F_{04}^3)^2
  \left[G^{22}G^{33}(\check F_{23}^2)^2(\sigma^3\sigma^2)^2-G^{00}G^{22}(\check
    F_{02}^3)^2(\sigma^3)^4-G^{22}G^{44}(\check F_{24}^3)^2(\sigma^3)^4\right]\\
  &+G^{00}G^{22}G^{33}G^{44}
  \left[(F_{34}^1)(F_{04}^3)(\check F_{02}^3)(\check
  F_{23}^2)\sigma^1\sigma^2(\sigma^3)^2-(F_{03}^2)(F_{04}^3)(\check
  F_{24}^3)(\check
  F_{23}^2)(\sigma^2\sigma^3)^2\right]\\
  &-G^{00}G^{22}G^{33}G^{44}(F_{03}^2)(F_{34}^1)(\check
  F_{02}^3)(\check F_{24}^3)\sigma^1\sigma^2(\sigma^3)^2\Bigg]\,.
\end{split}
\end{equation}\normalsize
Using the results for the symmetrized traces \eqref{eq:str}, the effective
Lagrangian simplifies to
\begin{equation}
  \label{eq:effLa32simpl}
  \begin{split}
    \call_{\text{eff}}^{a_2^3}=&\sqrt{-G}\:\Bigg[\frac{1}{2}G^{22}G^{44}(\check
    F_{24})^2+\frac{1}{2}G^{00}G^{22}(\check
    F_{02}^3)^2+\frac{1}{2}G^{22}G^{33}(\check F_{23}^2)^2\\
    &+\frac{1}{6}G^{33}G^{44}(F_{34}^1)^2
    \left[G^{00}G^{22}(\check F_{02}^3)^2-G^{22}G^{44}(\check
      F_{24}^3)^2-G^{22}G^{33}(\check F_{23}^2)^2\right]\\
    &+\frac{1}{2}G^{00}G^{44}(F_{04}^3)^2
    \left[\frac{1}{3}G^{22}G^{33}(\check
      F_{23}^2)^2-\frac{1}{2}G^{22}G^{44}(\check
      F_{24}^3)^2-\frac{1}{2}G^{00}G^{22}(\check F_{02}^3)^2\right]\\
    &+\frac{1}{2}G^{00}G^{33}(F_{03}^2)^2
    \left[-\frac{1}{3}G^{00}G^{22}(\check
      F_{02}^3)^2-\frac{1}{2}G^{22}G^{33}(\check
      F_{23}^2)^2+\frac{1}{3}G^{22}G^{44}(\check F_{24}^3)^2\right]\\
    &-\frac{2}{3}G^{00}G^{22}G^{33}G^{44}(F_{03}^2)(F_{04}^2)(\check
    F_{24}^3)(\check F_{23}^2)\Bigg]\,.
  \end{split}
\end{equation}
To simplify the equation of motion for the fluctuation $a_2^3$ (see
\eqref{eq:eoma32exp}) we define 
\begin{equation}
  \label{eq:4}
  \begin{split}
    \H^{00}&=G^{00}
    \left[1+\frac{1}{3}G^{33}G^{44}(\del_\rho\At_3^1)^2-\frac{1}{2}G^{00}G^{44}(\del_\rho\At_0^3)^2-\frac{1}{3}\frac{\vrho_H^4}{R^4}G^{00}G^{33}
      \left(\frac{\gamma}{\sqrt{2}}\At_3^1\At_0^3\right)^2\right]\,,\\
    \H^{33}&=G^{33}
    \left[1-\frac{1}{3}G^{33}G^{44}(\del_\rho\At_3^1)^2-\frac{1}{3}G^{00}G^{44}(\del_\rho\At_0^3)^2-\frac{1}{2}\frac{\vrho_H^4}{R^4}G^{00}G^{33}
      \left(\frac{\gamma}{\sqrt{2}}\At_3^1\At_0^3\right)^2\right]\,,\\
    \H^{44}&=G^{44}
    \left[1-\frac{1}{3}G^{33}G^{44}(\del_\rho\At_3^1)^2-\frac{1}{2}G^{00}G^{44}(\del_\rho\At_0^3)^2+\frac{1}{3}\frac{\vrho_H^4}{R^4}G^{00}G^{33}
      \left(\frac{\gamma}{\sqrt{2}}\At_3^1\At_0^3\right)^2\right]\,.
  \end{split}
\end{equation}

\end{appendix}


\providecommand{\href}[2]{#2}\begingroup\raggedright\endgroup

\bibliographystyle{felice_utcaps}

\end{document}